\documentclass[aps,prd,preprint,amsmath,amssymb]{revtex4}
\usepackage{epsfig,float}
\usepackage{graphicx}
\usepackage{dcolumn}
\usepackage{bm}
\usepackage{morefloats}

\newcommand{\be}{\begin{equation*}}
\newcommand{\ee}{\end{equation*}}
\newcommand{\bea}{\begin{eqnarray}}
\newcommand{\eea}{\end{eqnarray}}
\newcommand{\bean}{\begin{eqnarray*}}
\newcommand{\eean}{\end{eqnarray*}}

\begin{document}

\title{Dynamically generated $N^{*}$ and $\Lambda^*$ resonances in the hidden charm sector around 4.3 GeV}

\author{Jia-Jun Wu$^{1,2}$, R.~Molina$^{2,3}$, E.~Oset$^{2,3}$ and B.~S.~Zou$^{1,3}$ \\
$^1$ Institute of High Energy Physics, CAS, P.O.Box 918(4), Beijing 100049, China\\
$^2$ Departamento de F\'{\i}sica Te\'orica and IFIC, Centro Mixto
Universidad de Valencia-CSIC, Institutos de Investigaci\'on de
Paterna, Aptdo. 22085, 46071 Valencia, Spain\\
$^3$ Theoretical Physics Center for Science Facilities, CAS,
Beijing 100049, China}

\date{Feb. 2, 2011}

\begin{abstract}
The interactions of $\bar{D}\Sigma_{c}$-$\bar D\Lambda_c$,
$\bar{D}^{*}\Sigma_{c}$-$\bar D^*\Lambda_c$, and related
strangeness channels, are studied within the framework of the
coupled channel unitary approach with the local hidden gauge
formalism. A series of meson-baryon dynamically generated
relatively narrow $N^*$ and $\Lambda^*$ resonances are predicted
around 4.3 GeV in the hidden charm sector. We make estimates of
production cross sections of these predicted resonances in
$\bar{p} p $ collisions for PANDA at the forthcoming FAIR
facility.
\end{abstract}


\maketitle

\section{Introduction}
\label{s1}

The use of chiral Lagrangians in combination with unitary techniques
in coupled channels of mesons and baryons has been a very fruitful
scheme to study the nature of many hadron resonances. The poles
found in the analysis of meson baryon scattering amplitudes are
identified with existing baryon resonances. In this way the
interaction of the octet of pseudoscalar mesons with the octet of
stable baryons has lead to $J/P = 1/2^{-}$ resonances which fit
quite well the spectrum of the known low lying resonances with these
quantum
number~\cite{Kaiser:1995eg,angels,ollerulf,carmenjuan,hyodo}. The
combination of pseudoscalars with the decuplet of baryons has also
received attention and also leads to several dynamically generated
states~\cite{kolodecu,Sarkar:2004jh}. Work substituting pseudoscalar
mesons with vector mesons has also been done recently leading to new
resonances dynamically generated~\cite{sourav,angelsvec}.

One of the interesting findings in the study of the interaction of
pseudoscalars with the octet of baryons is the generation of the
$N^*(1535)$ resonance which has large couplings to $K \Sigma$ and $K
\Lambda$, to the point that the resonance can be approximately
considered as a bound state of these meson baryon
components~\cite{siegel,inoue,Nievesar}. Another point of view is
that this resonance can be considered as a hidden strangeness state.
In fact, phenomenological studies show that, indeed, this seems to
be the case~\cite{Liu:2005pm,Geng:2008cv}.

The idea that we want to explore here is to see if one can also
generate dynamically baryon states in the hidden charm sector. The
interaction of charmed mesons with the octet of stable baryons has
been studied in \cite{Lutz:2005vx,Hofmann:2005sw} and further
refined in \cite{Mizutani:2006vq,Tolos:2007vh,GarciaRecio:2008dp}.
Several states with open charm are dynamically generated there, in
particular the $\Lambda_c(2593)$.

In the present work we follow the steps of
\cite{Mizutani:2006vq,angelsvec} but concentrate in states of
hidden charm, for which we study the interaction of an anticharmed
meson with a charmed baryon. The underlying theory that we use is
an extension to SU(4) of the local hidden gauge Lagrangians
\cite{hidden1,hidden2,hidden3,hidden4}, where SU(4) is broken to
account for the different masses of the vector mesons exchanged in
the t- and u- channels. The study is done both with pseudoscalar
mesons and vector mesons and we obtain three dynamically generated
hidden charm baryons generated from the pseudoscalar baryon
interaction plus three other states from the interaction of vector
mesons with baryons, all of them with masses around 4200-4600 MeV.

We also make estimates of production cross sections with $\bar{p}$
collisions that could be carried out at the future FAIR facility
within the PANDA project. We also study how the presence of these
resonances could increase the rate of $J/\psi$ and $\eta_c$
production around the energies where the resonances can be formed.
Part of our results have been briefly reported in \cite{wu-prl},
here we give a much more complete report of our investigation.

In the next section, we present the formalism and ingredients for
the study of the interaction, and give the poles obtained. In the
last section, our numerical results are given, followed by a
discussion.

\section{Formalism for Meson-Baryon Interaction}
\label{s2}
\subsection{Lagrangian and Feynman diagrams }
\begin{figure}[htbp] \vspace{-0.cm}
\begin{center}
\includegraphics[width=0.7\columnwidth]{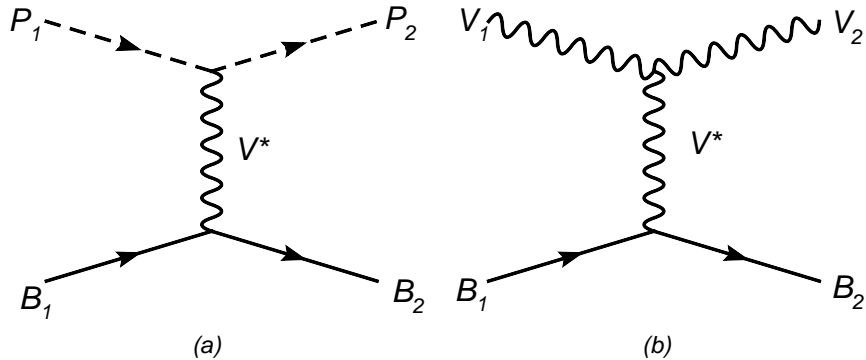}
\caption{Feynman diagrams for the pseudoscalar-baryon (a) or
vector- baryon (b) interaction via the exchange of a vector meson
($P_{1}$, $P_{2}$ are $D^{-}$, $\bar{D}^{0}$ or $D^{-}_{s}$, and
$V_{1}$, $V_{2}$ are $D^{*-}$, $\bar{D}^{*0}$ or $D^{*-}_{s}$, and
$B_{1}$, $B_{2}$ are $\Sigma_{c}$, $\Lambda^{+}_{c}$, $\Xi_{c}$,
$\Xi'_{c}$ or $\Omega_{c}$, and $V^{*}$ is $\rho$, $K^{*}$, $\phi$
or $\omega$).} \label{fe}
\end{center}
\end{figure}

We consider the $PB\rightarrow PB$ and $VB\rightarrow VB$
interaction by exchanging a vector meson. The corresponding
Feynman diagrams are shown in the Fig.\ref{fe}.

In order to evaluate these Feynman diagrams, we give the three
types of vertices for BBV, PPV and VVV interaction from
\cite{angelsvec}. The Lagrangians for the interaction of vector
mesons between themselves (three - vector vertex), pseudoscalar
mesons with vectors and baryons with vectors are:
\begin{eqnarray}
{\cal L}_{VVV}&=&ig\langle V^\mu[V^{\nu},\partial_\mu V_{\nu}]\rangle\nonumber\\
{\cal L}_{PPV}&=&-ig\langle V^\mu[P,\partial_\mu P]\rangle\nonumber\\
{\cal L}_{BBV}&=&g (\langle\bar{B}\gamma_\mu
[V^\mu,B]\rangle+\langle\bar{B}\gamma_\mu B\rangle\langle
V^\mu\rangle)\ \label{eq:lag}
\end{eqnarray}
where $B$ and $P$ are the standard matrices including the
pseudoscalar and baryon nonets \cite{sunra}, $g=M_V/2 f$ is the
coupling used in the hidden gauge with the pion decay constant
$f=93$ MeV and the mass of the light vector meson taken as $M_V=770$
MeV. The $g$ fulfills the KSFR rule \cite{ksfr} which is tied to
vector meson dominance \cite{vector}. When we go to SU(4) we can
still use the Lagrangian for VPP of Eq. (\ref{eq:lag}) and the $V$
and $P$ matrices extended to SU(4):
\begin{equation}
\renewcommand{\tabcolsep}{1cm}
\renewcommand{\arraystretch}{2}
P=\left(
\begin{array}{cccc}
\frac{\pi^0}{\sqrt{2}}+\frac{\eta_8}{\sqrt{6}} +\frac{\tilde{\eta}_c}{\sqrt{12}}+\frac{\tilde{\eta}_c'}{\sqrt{4}}  &\pi^+     & K^{+}    &\bar{D}^{0}\\
\pi^-      & -\frac{\pi^0}{\sqrt{2}}+\frac{\eta_8}{\sqrt{6}}+\frac{\tilde{\eta}_c}{\sqrt{12}}+\frac{\tilde{\eta}_c'}{\sqrt{4}}& K^{0}    &D^{-}\\
K^{-}      & \bar{K}^{0}       &\frac{-2\eta_8}{\sqrt{6}}+\frac{\tilde{\eta}_c}{\sqrt{12}}+\frac{\tilde{\eta}_c'}{\sqrt{4}}                            &D^{-}_s\\
D^{0}&D^{+}&D^{+}_s&-\frac{3\tilde{\eta}_c}{\sqrt{12}}+\frac{\tilde{\eta}_c'}{\sqrt{4}}\\
\end{array}
\right), \label{eq:P}
\end{equation}

and
\begin{equation}
\renewcommand{\tabcolsep}{1cm}
\renewcommand{\arraystretch}{2}
V_\mu=\left(
\begin{array}{cccc}
\frac{\rho^0}{\sqrt{2}}+\frac{\omega_8}{\sqrt{6}} +\frac{\tilde{\omega}_c}{\sqrt{12}}+\frac{\tilde{\omega}_c'}{\sqrt{4}}& \rho^+ & K^{*+}&\bar{D}^{*0}\\
\rho^- &-\frac{\rho^0}{\sqrt{2}}+\frac{\omega_8}{\sqrt{6}}+\frac{\tilde{\omega}_c}{\sqrt{12}}+\frac{\tilde{\omega}_c'}{\sqrt{4}} & K^{*0}&D^{*-}\\
K^{*-} & \bar{K}^{*0} &\frac{-2\omega_8}{\sqrt{6}}+\frac{\tilde{\omega}_c}{\sqrt{12}}+\frac{\tilde{\omega}_c'}{\sqrt{4}}&D^{*-}_s\\
D^{*0}&D^{*+}&D^{*+}_s&-\frac{3\tilde{\omega}_c}{\sqrt{12}}+\frac{\tilde{\omega}_c'}{\sqrt{4}}\\
\end{array}
\right)_\mu\ . \label{eq:V}
\end{equation}

Let us recall that here $\tilde{\eta}_c$ stands for the SU(3)
singlet of the 15th SU(4) representation and we denote
$\tilde{\eta}'_c$ for the singlet of SU(4). On the other hand,
$\omega_8$ plays the role of the $\eta_8$ for the vectors, while
$\tilde{\omega}_c$ the role of $\tilde{\eta}_c$, and we denote by
$\tilde{\omega}'_c$ the SU(4) singlet. We take $\pi^0$, $\eta$,
$\eta'$, $\eta_c$ as a basis for the neutral pseudoscalar mesons,
where $\eta'$ is the singlet in SU(3), $(u\bar{u}+ d\bar{d}+
s\bar{s})/\sqrt{3}$, and $\eta_c$ stand for $c\bar{c}$. Recalling
the standard quark composition of the SU(4) mesons
\begin{eqnarray}
 \pi^0&=&\frac{1}{\sqrt{2}}(u\bar{u}-d\bar{d})\nonumber\\
\eta_8&=&\frac{1}{\sqrt{6}}(u\bar{u}+d\bar{d}-2 s\bar{s})\nonumber\\
\tilde{\eta}_c&=&\frac{1}{\sqrt{12}}(u\bar{u}+d\bar{d}+ s\bar{s}-3c\bar{c})\nonumber\\
\tilde{\eta}'_c&=&\frac{1}{\sqrt{4}}(u\bar{u}+d\bar{d}+
s\bar{s}+c\bar{c})\ , \label{eq:mat}
\end{eqnarray}
we find
\begin{eqnarray}
\eta_{8}&=&\eta\nonumber\\
\eta'&=&\frac{1}{2} \tilde{\eta}_c+\frac{\sqrt{3}}{2}\tilde{\eta}'_c\nonumber\\
\eta_c&=&\frac{1}{2}(-\sqrt{3}\tilde{\eta}_c+\tilde{\eta}'_c)\ ,
\label{eq:ph}
\end{eqnarray}
in the physical basis. On the other hand, for vectors we use the
physical basis $\rho^0$, $\omega$, $\phi$ and $J/\psi$, where
\begin{eqnarray}
 \rho^0&=& \frac{1}{\sqrt{2}}(u\bar{u}-d\bar{d})\nonumber\\
\omega&=& \frac{1}{\sqrt{2}}(u\bar{u}+d\bar{d})\nonumber\\
\phi&=& s\bar{s}\nonumber\\
J/\psi&=& c\bar{c}\ , \label{eq:mat1}
\end{eqnarray}
which can be written in terms of $\omega_8$, $\tilde{\omega}_c$
and $\tilde{\omega}_c'$ as$^1$ \footnotetext[1]{Latter on, in
order to use the SU(4) Clebsch Gordan coefficients we shall change
a phase to the $\tilde{\eta}_c$ and $\tilde{\omega}_c$.}
\begin{eqnarray}
 \omega&=&\frac{1}{6}(\sqrt{6}\tilde{\omega}_c+2\sqrt{3}\omega_8+3\sqrt{2}\tilde{\omega}'_c)\nonumber\\
\phi&=&\frac{1}{6}(\sqrt{3}\tilde{\omega}_c-2\sqrt{6}\omega_8+3\tilde{\omega}'_c)\nonumber\\
J/\psi&=&\frac{1}{2}(-\sqrt{3}\tilde{\omega}_c+\tilde{\omega}'_c)\
. \label{eq:ph1}
\end{eqnarray}
 The use of
Lagrangians to give the BBV vertex in SU(4) is more cumbersome
than in SU(3) and thus it is simpler to use SU(4) Clebsch Gordan
coefficients. Yet, this requires a certain phase convention for
the physical states with respect to the isospin states implicit in
the SU(4) tables, which makes convenient to use the same procedure
to evaluate the $PPV$ vertex.

In the $PPV$ vertex we go from the $15\otimes15$ representation of
pseudoscalars to the $15$ representation of vectors. Yet, the
nature of the couplings (with the explicit commutator) has as a
consequence that only the $15_F$ (antisymmetric) representation
for the vectors is needed (one can alternatively use explicitely the $15_F$ and the $15_D$ representations and the $15_D$ contribution vanishes at the end$^2$\footnotetext[2]{We thank J. Nieves for pointing this to us.}). The resulting $t$ amplitude for
$P_1P_2\to V$ is given by
\begin{equation}
 t_{P_1P_2V}=g_{15_F}\, C_{15_F}(15\otimes15)\,(q_1+q_2)\cdot\epsilon\ ,
\label{eq:mes}
\end{equation}
where $q_1$ and $q_2$ are the four-momentum of the initial and
final pseudoscalar mesons respectively, and
$C_{15_F}(15\otimes15)$ is the SU(4) Clebsch Gordan Coefficient
that we take from \cite{Haacke} and $g_{15_F}$ is the reduced
matrix element that by comparison with the result of the
Lagrangian is given by
\begin{equation}
 g_{15_F}=-2\sqrt{2} g \ .
\label{eq:g}
\end{equation}
However, the use of the SU(4) tables requires a phase convention. We
find a compatible and convenient phase convention of the isospin
states implicit in the SU(4) tables and those used by us in Eqs.
(\ref{eq:mes}) and (\ref{eq:g}) by means of:
\begin{center}
\begin{tabular}{lll}
$|K^0\rangle=-|1/2,-1/2\rangle\ ;$&$|\pi^+\rangle=-|1,1\rangle\ ;$&$|\pi^0\rangle=-|1,0\rangle\ ;$\\\\
$|D^+_s\rangle=-|0,0\rangle\ ;$&$|\bar{D}^0\rangle=-|1/2,1/2\rangle\ ;
$&$|\tilde{\eta}_c\rangle=-|0,0\rangle\ ;$\\
\end{tabular}
\end{center}
and equivalent phases for the corresponding vectors, $K^{*0}$,
$\rho^+$, $\rho^0$, $D^{*+}_s$, $\bar{D}^{*0}$ and
$\tilde{\omega}_c$. The necessity for the change in phases stems
from demanding that the $15\otimes 15\to1$ combination of SU(4)
isospin states is a symmetrical expression in the physical states
\cite{swartz}. The use of this convention (and also the convention
for baryons that we give later) leads to the same amplitudes in
charge basis given by the Lagrangians of Eq. (\ref{eq:lag}) with
the $P$ and $B$ matrices written in the SU(3) basis.

When we go to the $BBV$ vertex (we look for $B\bar{B}\to V$), we
need now the three representations, $15_1$, $15_2$ and $1$, and we
must note that when the $8$ representation of SU(3) is involved,
only the $F$ coefficients are needed. In this case we have
$20'\otimes \bar{20}'\to 15_1,15_2,1$, and the $t$ amplitude for
the BBV vertex is given by
\begin{eqnarray}
t_{B_1\bar{B}_2V} =\lbrace g_{15_1}\,C_{15_1}(20'\otimes\bar{20}')
+g_{15_2}\,C_{15_2}(20'\otimes\bar{20}')+g_{1}\,C_{1}(20'\otimes\bar{20}')
\rbrace\,\bar{u}_{r'}(p_2)\gamma\cdot\epsilon\, u_r(p_1)\ .\nonumber\\
\label{eq:bbv}
\end{eqnarray}
Once again by writing the expression for $20'\otimes\bar{20}'\to1$
in terms of the SU(4) isospin states, and demanding that the
expression is symmetrical in the physical baryons, we obtain a
convention of phases. The one we have chosen, partly motivated to
agree formally with earlier SU(3) results, is given by changing
the phases of the states
\begin{center}
 \begin{tabular}{llll}
$|\bar{\Xi}^{--}_{cc}\rangle=-|1/2,-1/2\rangle\ ,$  & $|\Omega^+_{cc}\rangle=-|0,0\rangle\ ,$        & $|\Xi^0_c\rangle=-|1/2,-1/2\rangle\ ,$         &  $|\Xi'^0_c\rangle=-|1/2,-1/2\rangle\ ,$\\\\
$|\bar{\Lambda}^-_c\rangle=-|0,0\rangle\ ,$         &$|\Sigma_{c}^+\rangle=-|1,0\rangle\ ,$          &$|\Sigma_{c}^{++}\rangle=-|1,1\rangle\ ,$       &$|\bar{\Sigma}_{c}^{--}\rangle=-|1,-1\rangle\ ,$\\\\
$|n\rangle=-|1/2,-1/2\rangle\ ,$                    & $|\bar{\Xi}^0\rangle=-|1/2,-1/2\rangle\ ,$     & $|\bar{\Sigma}^+\rangle=-|1,1\rangle\ ,$       &  $|\Sigma^+\rangle=-|1,1\rangle\ ,$\\\\
$|\Sigma^0\rangle=-|1,0\rangle\ ,$
&$|\bar{\Sigma}^{0}\rangle=-|1,0\rangle\ $ .
 \end{tabular}
\end{center}
The reduced matrix elements of Eq. (\ref{eq:bbv}), $g_{15_1}$,
$g_{15_2}$ and $g_1$ are evaluated demanding:
\begin{itemize}
\item[1)] The coupling $p\bar{p}\to J/\psi$ should be zero by OZI
rules, \item[2)] The coupling $p\bar{p}\to \phi$ should be zero by
OZI rules, \item[3)] The coupling $p\bar{p}\to \rho^0$ should be
the one obtained in SU(3).
\end{itemize}
We finally obtain
\begin{eqnarray}
 g_{15_1}=-g;\hspace{0.5cm}g_{15_2}=2\sqrt{3}\,g;\hspace{0.5cm}g_{1}=3\sqrt{5}\,g\ .
\end{eqnarray}
with $g=M_V/2f$ and $f = 93 MeV$ the pion decay constant.

The diagram of Fig. \ref{fe} (a) requires the exchange of the
vector meson with the two vertices given by Eqs. (\ref{eq:mes})
and (\ref{eq:bbv}). In the sum of polarizations in the vector
meson exchanged,
\begin{equation}
\sum_\lambda \epsilon_\mu \epsilon_\nu =-g_{\mu\nu}+\frac{q_\mu q_\nu}{M^2_V}\ .
\end{equation}
We can keep just the $\mu=\nu=0$ component since we assume that
the three momenta of the particles are small compared to their
masses. Similarly, the $q^2/M^2_V$ term in the vector meson
propagator is neglected (further on, when we consider the
transitions from heavy mesons to light ones, we perform the exact
calculation). The transition potential corresponding to the
diagrams of Fig. \ref{fe} are given by
\begin{eqnarray}
V_{ab(P_{1}B_{1}\rightarrow P_{2}B_{2})}&=&\frac{C_{ab}}{4f^{2}}(q^0_1+q^0_2)\label{vpbb},\\
V_{ab(V_{1}B_{1}\rightarrow
V_{2}B_{2})}&=&\frac{C_{ab}}{4f^{2}}(q^0_1+q^0_2)\vec{\epsilon}_1\cdot\vec{\epsilon}_2\ .\label{vvbb}
\end{eqnarray}
Where the indices $a,b$ stand for different groups of $P_{1}(V_{1})B_{1}$
and $P_{2}(V_{2})B_{2}$, respectively. The $q^0_1,q^0_2$ are the
energies of the initial, final meson. We list the value of the
$C_{ab}$ coefficients for different states of isospin, I, and
strangeness, S in the Appendix. Here we study six different cases
with $(I, S) = (3/2, 0), (1/2, 0), (1/2, -2),(1, -1),(0, -1),(0,
-3)$.

\subsection{The $G$ function and the unitary $T$ amplitudes}
The $G$ function is a loop function of a meson ($P$) and a
baryon($B$) which we calculate in dimensional regularization by
means of the formula

\begin{eqnarray}
G_{(P,B)}&=&i2M_{B}\int\frac{d^{4}q}{(2\pi)^{4}}\frac{1}{(P-q)^{2}-M^{2}_{B}+i\varepsilon}\,\frac{1}{q^{2}-M^{2}_{P}+i\varepsilon},\\
&=&\frac{2M_{B}}{16\pi^2}\big\{a_{\mu}+\textmd{ln}\frac{M^{2}_{B}}{\mu^{2}}+\frac{M^{2}_{P}-M^{2}_{B}+s}{2s}\textmd{ln}\frac{M^{2}_{P}}{M^{2}_{B}}\nonumber\\
&&+\frac{\bar{q}}{\sqrt{s}}\big[\textmd{ln}(s-(M^{2}_{B}-M^{2}_{P})+2\bar{q}\sqrt{s})+\textmd{ln}(s+(M^{2}_{B}-M^{2}_{P})+2\bar{q}\sqrt{s})\nonumber\\
&&-\textmd{ln}(-s-(M^{2}_{B}-M^{2}_{P})+2\bar{q}\sqrt{s})-\textmd{ln}(-s+(M^{2}_{B}-M^{2}_{P})+2\bar{q}\sqrt{s})\big]\big\}\ ,\label{Gf}
\end{eqnarray}
where
\begin{eqnarray}
s&=&P^2,\\
\bar{q}&=&\frac{\sqrt{(s-(M_{B}+M_{P})^2)(s-(M_{B}-M_{P})^2)}}{2\sqrt{s}}\hspace{0.2cm}\mathrm{with}\hspace{0.2cm}\mathrm{Im}(q)>0\ .
\end{eqnarray}
In Eq. (\ref{Gf}), $q$ is the four-momentum of the meson, and $P$
is the total four-momentum of the meson and the baryon. The $\mu$
is a regularization scale, which we put 1000 MeV, and $a_{\mu}$ is
of the order of $-2$, which is the natural value of the
subtraction constant \cite{ollerulf}. When we look for poles in
the second Riemann sheet, we must change $q$ by $-q$ when
$\mathrm{\sqrt{s}}$ is above the threshold in Eq. (\ref{Gf})
\cite{luisaxial}. See further comments regarding the subtraction constant in Subsection D.

Here we also regularize the $G$ loop function in a different way
by putting a cutoff in the three-momentum. The formula is:
\begin{eqnarray}
G_{(P,B)}&=&i2M_{B}\int\frac{d^{4}q}{(2\pi)^{4}}\frac{1}{(P-q)^{2}-M^{2}_{B}+i\varepsilon}\frac{1}{q^{2}-M^{2}_{P}+i\varepsilon}\nonumber\\
&=&\int^{\Lambda}_{0}\frac{q^{2}dq}{4\pi^{2}}\frac{2M_{B}(\omega_{P}+\omega_{B})}{\omega_{P}\,\omega_{B}\,((P^0)^{2}-(\omega_{P}+\omega_{B})^{2}+i\epsilon)}\ ,\label{Gf2}
\end{eqnarray}
where
\begin{eqnarray}
\omega_{P}&=&\sqrt{\vec{q}\,^{2}+M^{2}_{P}}\ ,\nonumber\\
\omega_{B}&=&\sqrt{\vec{p}\,^{2}+M^{2}_{B}}\ ,
\end{eqnarray}
and $\Lambda$ is the cutoff parameter in the three-momentum of the
function loop.

For these two types of $G$ function, the free parameters are
$a_{\mu}$ in Eq. (\ref{Gf}) and $\Lambda$ in Eq. (\ref{Gf2}). When
we choose $a_\mu$ or $\Lambda$, the shapes of these two functions
are almost the same close to threshold and they take the same
value at threshold.

Then we can get the unitary T amplitudes by solving the coupled
channels Bethe-Salpeter equation in the on shell factorization
approach of \cite{nsd,ollerulf,Nieves:1999bx}
\begin{eqnarray}
T=[1-VG]^{-1}V\ .\label{Bethe}
\end{eqnarray}

When we look for poles in the complex plane of $\sqrt{s}$, poles
in the $T$ matrix that appear in the first Riemann sheet below
threshold are considered as bound states whereas those located in
the second Riemann sheet and above the threshold of some channel
are identified as resonances.

\subsection{The coupling constant and the width of the poles}

From the $T$ matrix we can find the pole positions $z_{R}$. In
this work, we find all of these poles in the real axes below
threshold, in a few words, they are bound states. In view of that,
for these cases the coupling constants are obtained from the
amplitudes in the real axis. These amplitudes behave close to the
pole as:
\begin{eqnarray}
T_{ab}=\frac{g_{a}g_{b}}{\sqrt{s}-z_{R}}\ .
\end{eqnarray}
We can get the coupling constant as:
\begin{eqnarray}
g_{a}^{2}=\lim_{\sqrt{s}\rightarrow
z_{R}}(T_{aa}\times(\sqrt{s}-z_{R})).\label{coupling1}
\end{eqnarray}
This expression allows us to determine the value of $g_{a}$,
except by a global phase. Then, the other couplings are derived
from
\begin{eqnarray}
g_{b}=\lim_{\sqrt{s}\rightarrow
z_{R}}(\frac{g_{a}T_{ab}}{T_{aa}})\ .\label{coupling2}
\end{eqnarray}

\begin{figure}[htbp] \vspace{-0.cm}
\begin{center}
\includegraphics[width=0.7\columnwidth]{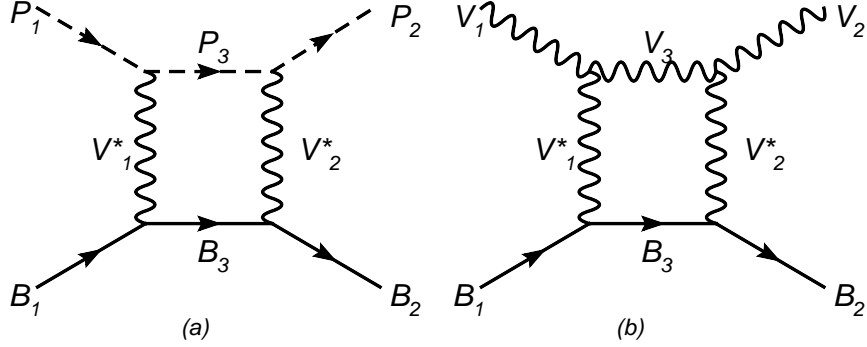}
\caption{The Feynman diagrams of pseudoscalar-baryon (a) or
vector-baryon (b) interaction via a box diagram. $P_{1}$, $P_{2}$,
$V_{1}$, $V_{2}$, $B_{1}$, $B_{2}$ are the same particles than in
Fig. \ref{fe}. $P_{3}$, $V_{3}$ and $B_{3}$ are light particles
belonging to the SU(3) octet of pseudoscalar mesons, vector mesons
and stable baryons, respectively, and $V^{*}_{1}$, $V^{*}_{2}$ are
$D^{*}$ or $D^{*}_{s}$.} \label{wfe}
\end{center}
\end{figure}

As all the states that we find have zero width, we should take
into account some decay mechanisms. Thus, we consider the decay of
the states to light meson - light baryon by means of box diagrams
as it was done in \cite{raquel,geng}. The Feynman diagrams for
these decays are shown in Fig. \ref{wfe}. We assume that $P_{3}$,
$V_{3}$ and $B_{3}$ are on-shell and neglect the three - momentum
of the initial and final particles. Then, using Eq.
(\ref{eq:lag}), the transition potential of these diagrams can be
written as:
\begin{eqnarray}
V_{acb(P_{1}B_{1}\rightarrow P_{3}B_{3}\rightarrow
P_{2}B_{2})}&=&\frac{C_{ac}C_{cb}M^{4}_{V^{*}}}{16f^{4}}\times G_{(P_{3},B_{3})}\times\frac{(\sqrt{s}+M_{B_3})^2-M_{P_3}^2}{4\sqrt{s} \, M_{B_3}}\nonumber\\
&&\times\frac{-2E_{P_{1}}+(M_{B_{3}}-M_{B_{1}})(M^2_{P_{1}}+M^2_{V^*_{1}}-M^2_{P_{3}})/M^{2}_{V^{*}_{1}}}
{M^2_{P_{1}}+M^2_{P_{3}}-2E_{P_{3}}E_{P_{1}}-M^2_{V^*_{1}}}\nonumber\\
&&\times\frac{-2E_{P_{2}}+(M_{B_{3}}-M_{B_{2}})(M^2_{P_{2}}+M^2_{V^*_{2}}-M^2_{P_{3}})/M^{2}_{V^{*}_{2}}}
{M^2_{P_{2}}+M^2_{P_{3}}-2E_{P_{3}}E_{P_{2}}-M^2_{V^*_{2}}},\label{widpb}
\end{eqnarray}
and the same for vectors (see Fig. \ref{wfe}. (b)) changing
$E_{P_1}$, $E_{P_2}$, $E_{P_3}$ by $E_{V_1}$, $E_{V_2}$, $E_{V_3}$
and $M_{P_1}$, $M_{P_2}$, $M_{P_3}$ by $M_{V_1}$, $M_{V_2}$,
$M_{V_3}$, respectively. Here $c$ stands for a different group of
$P_{3}(V_{3})B_{3}$. Then, the kernel $V$ in the Bethe Salpeter
equation, Eq. (\ref{Bethe}), becomes now:
\begin{eqnarray}
V_{ab(P_{1}B_{1}\rightarrow
P_{2}B_{2})}&=&\frac{C_{ab}}{4f^{2}}(E_{P_{1}}+E_{P_{2}})+\sum_{c}V_{acb}\
,
\end{eqnarray}
and similarly for the $VB$ system. In Eq. (\ref{widpb}) we have
factorized the two $P_1 B_1\to P_3 B_3$ and $P_3 B_3\to P_2 B_2$
transition amplitudes outside the loop integral by taking their
values when the system $P_3 B_3$ is set on-shell. This is a good
approximation, exact for the imaginary part of the diagram, which is
our main concern, since we are interested in the contribution of
these diagrams to the width of the resonances. The loop integral
only affects then the $P_3$, $B_3$ propagators leading to the same
$G$ function defined in Eq. (\ref{Gf}). Since the on-shell mass of
the intermediate states is far away from the energies investigated,
$\mathrm{Re}G(P_3,B_3)$ is small and we have checked that it is
sufficiently smaller than the tree level contribution from the
diagrams of Fig. (\ref{fe}), such that it can be ignored. For
example, $V_{(\bar{D}\Sigma_c\to \pi N\to
\bar{D}\Sigma_c)}=(0.38+2.9i)$~GeV$^{-1}$ at the $N^*$ pole position
with $\sqrt{s}=4.265$ GeV.

Further on, we will include the $\eta_c N$, $\eta_c\Lambda$
channels for $PB\to PB$, and $J/\psi N$, $J/\psi \Lambda$ for
$VB\to VB$ in the calculation.
\subsection{Discussion about the use of the SU(4) flavor symmetric Lagrangian}
Once the formalism has been exposed we would like to make some comments to
justify the approach. While the hidden local gauge approach is well settled with
SU(3) flavor, its extension to SU(4) is not quite justified. The local hidden
gauge theory in QCD is based on spontaneous symmetry breaking of chiral
symmetry, which is not expected to hold in the charmed sector, and even if this
were the case, the breaking pattern would be masked by the large charm mass. In
view of this, the approach followed requires some justification. The first thing
we must bare in mind is that the large mass of the charmed quark is to be blamed
for the  lack of symmetry. Hence, if hadron dynamics still has some traces of SU(4)
symmetry it should be in particular vertices or amplitudes
not tied to the quark mass. In this respect it is interesting to mention that SU(4) symmetry
works fairly well for the vertices VP$\gamma$ (or equivalently the VVP) involved in the radiative decay of
vector mesons with charm to a pseudoscalar and a photon \cite{tesisdaniel}. The agreement is as good as in
SU(3). Given the analogies between the VVP and VVV vertices provided by heavy quark symmetry, it is fair to think that using SU(4) symmetry to evaluate the VVV and VPP vertices would be a fair starting point.
 Similarly we could also assume the same symmetry
to hold in the VBB vertices.

 One should also note that in the case of meson meson interaction the present
approach provides the same results, up to a mass term of no
practical consequences, as the heavy quark formalism used in
\cite{kolo,hofmann,guo} for the case of interaction of light mesons
with heavy meson. From the perspective of our approach, we could
rephrase it by stating that these two approaches provide the same
VVV (or VPP) vertices with two heavy vectors and a light one. For
the evaluation of the width in the present approach one is using
these vertices, with the only difference with respect to
\cite{kolo,hofmann,guo} that one of the heavy vectors is exchanged
in the $t-$channel in our approach, while the two heavy vectors (or
pseudoscalars) were external particles in \cite{kolo,hofmann,guo}.
Only vertices involving three heavy vectors ($D^* D^* J/\psi$) would
require the extra help of SU(4) symmetry. Thus, most of the
information used is supported by phenomenology and other approaches.

Furthermore, we should bare in mind that the largest fraction of the
results that we obtain, concerning couplings to different channels,
is based on SU(3), since we can relate these channels through SU(3).
Then, implementing SU(4) symmetry as we do, automatically
accomplishes this. Only when we give a jump to another charm sector
the SU(4) symmetry would play a role and there we would invoke the
arguments used above to support it. Certainly, when these vertices
are used in Feynmann diagrams and the masses of the exchanged
vectors are very different, the approximate SU(4) symmetry that we
had in the vertices will be badly broken in the amplitudes. This
happens also in SU(3): The vertices are manifestly SU(3) invariant,
but when SU(3) is broken in the amplitudes because of the different
masses of the particles belonging to the same multiplet (for
instance in the unitarization procedure), the underlying SU(3)
symmetry is broken and two octets that were degenerate in the exact
SU(3) limit give rise to two different states in the strangeness
$S=-1$ sector: one of the two $\Lambda(1405)$ states and the
$\Lambda(1670)$.

Yet, one should be ready to accept larger uncertainties than in
SU(3) and allow some fitting freedom in the approach. This can be
done by means of the subtraction constants of the G function, that
effectively tune the strength of the potentials that one is using in
the approach. This also means that the natural values of these
constants should only be used as indicative and then a real fit to
the data should be done, which cannot be done in the present case
since we have no experimental data. However, one can rely on
previous work along these lines in which several groups have done
this work and provide the new scale of the subtraction constants to
be used in the charm sector. In this sense, the  works of
\cite{daniel1,daniel2,hideko,xyz,Mizutani:2006vq}, choosing these
parameters to reproduce properties of known resonances like the
$D^0_{s0}(2317)$, the X(3872), the $D_2^*(2460)$ and  $D^*(2640)$,
and $\Lambda_c(2593)$, have given us the scale for these
subtractions constants that we use here.

\section{Result and Discussion}

\subsection{The pole positions and coupling constants}

Here we show the results for the different sectors. By using the
two $G$ functions of Eqs. (\ref{Gf}) and (\ref{Gf2}), the poles
appear in both cases below threshold in the first Riemann sheet
and therefore they are bound states. We show the pole positions
for different values of $\alpha(\Lambda)$ in Tables \ref{pbpole}
and \ref{vbpole}.

We take a range of values of $\alpha$, or accordingly the cut off, in line with values used in \cite{daniel1,daniel2,Mizutani:2006vq,hideko} and we find six poles in our calculation. The uncertainties in the
pole positions in the case of the first and third poles for both
$PB$ and $VB$ systems, are of the order of $100$ MeV, which are
typical in any hadron model. These two poles are rather stable.
However, for the second state, the uncertainties are much larger
and the pole position is very unstable.
\begin{table}[ht]
      \renewcommand{\arraystretch}{1.1}
     \setlength{\tabcolsep}{0.4cm}
\begin{center}
\begin{tabular}{cccc}\hline
$(I, S)$&$\alpha=-2.2$($\Lambda=0.7$ GeV)  &  $\alpha=-2.3$($\Lambda=0.8$ GeV) & $\alpha=-2.4$($\Lambda=0.9$ GeV)\\
    &  $z_R$               &  $z_R$                &  $z_R$\\
\hline
$(1/2, 0)$    & $4291(4273)$             & $4269(4236)$                    & $4240(4187)$ \\
\hline
$(0, -1)$     & $4247(4120)$             & $4213(4023)$                    & $4170(3903)$  \\
              & $4422(4394)$             & $4403(4357)$                    & $4376(4308)$  \\
\hline\end{tabular} \caption{Pole position from $PB\rightarrow PB$
using the two different $G$ functions of Eqs. (\ref{Gf}) and
(\ref{Gf2}). The units are in MeV.}
 \label{pbpole}
\end{center}
      \renewcommand{\arraystretch}{1.1}
     \setlength{\tabcolsep}{0.4cm}
\begin{center}
\begin{tabular}{cccc}\hline
$(I, S)$&$\alpha=-2.2$($\Lambda=0.7$ GeV) &  $\alpha=-2.3$($\Lambda=0.8$ GeV) & $\alpha=-2.4$($\Lambda=0.9$ GeV)\\
    &  $z_R$               &  $z_R$                &  $z_R$\\
\hline
$(1/2, 0)$    & $4438(4410)$             & $4418(4372)$                   & $4391(4320)$ \\
\hline
$(0, -1)$     & $4399(4256)$             & $4370(4155)$                   & $4330(4030)$  \\
              & $4568(4532)$             & $4550(4493)$                      & $4526(4441)$  \\
\hline\end{tabular} \caption{Pole position from $VB\rightarrow VB$
using the two different $G$ functions of Eqs. (\ref{Gf}) and
(\ref{Gf2}). The units are in MeV.}
 \label{vbpole}
\end{center}
\end{table}

For the discussions that follow we choose an intermediate value of
$\alpha$, which we take $\alpha=-2.3$, to study the nature of
these poles in detail. In Tables \ref{pbcoupling} and
\ref{vbcoupling}, the values of the coupling constants are listed
by using Eqs. (\ref{coupling1}) and (\ref{coupling2}). From Table
\ref{pbcoupling}, we see that both the $N^{*}(4269)$ and the
$\Lambda^{*}(4403)$ depend on one channel, $\bar{D} \Sigma_{c}$
and $\bar{D} \Xi'_{c}$, respectively. These two states are both
stable as we can see in Table \ref{pbpole}. In contrast, the
$\Lambda^{*}(4213)$ depend on two channels, $\bar{D}_{s}
\Lambda^{+}_{c}$ and $\bar{D} \Xi_{c}$. The mass of this state
changes appreciably by using different values of the free
parameters ($\alpha$ or $\Lambda$).

\begin{table}[ht]
      \renewcommand{\arraystretch}{1.1}
     \setlength{\tabcolsep}{0.4cm}
\begin{center}
\begin{tabular}{ccccc}\hline
$(I, S)$ &  $z_R$ (MeV)    & \multicolumn{3}{c}{$g_a$}\\
\hline
$(1/2, 0)$    &      & $\bar{D} \Sigma_{c}$ & $\bar{D} \Lambda^{+}_{c}$ \\
          & $4269$ & $2.85$                 &  $0$\\
\hline
$(0, -1)$  &        & $\bar{D}_{s} \Lambda^{+}_{c}$   & $\bar{D} \Xi_{c}$ & $\bar{D} \Xi'_{c}$\\
       &   $4213$ & $1.37$                            & $3.25$              & $0$              \\
       &   $4403$ & $0$                               & $0$                 & $2.64$              \\
\hline\end{tabular} \caption{Pole positions, $z_R$ and coupling constants, $g_a$,
for the states from $PB\rightarrow PB$.}
 \label{pbcoupling}
\end{center}
      \renewcommand{\arraystretch}{1.1}
     \setlength{\tabcolsep}{0.4cm}
\begin{center}
\begin{tabular}{ccccc}\hline
$(I, S)$&  $z_R$ (MeV)   & \multicolumn{3}{c}{$g_a$}\\
\hline
$(1/2, 0)$    &      & $\bar{D}^{*} \Sigma_{c}$ & $\bar{D}^{*} \Lambda^{+}_{c}$ \\
          & $4418$ & $2.75$                     &  $0$\\
\hline
$(0, -1)$  &        & $\bar{D}^{*}_{s} \Lambda^{+}_{c}$   & $\bar{D}^{*} \Xi_{c}$ & $\bar{D}^{*} \Xi'_{c}$\\
       &   $4370$ & $1.23$                                & $3.14$                 & $0$                    \\
       &   $4550$ & $0$                                   & $0$                    & $2.53$                 \\
\hline\end{tabular} \caption{Pole position and coupling constants
for the bound states from $VB\rightarrow VB$.}
 \label{vbcoupling}
\end{center}
\end{table}

\subsection{The decay widths of these states to light meson - light baryon channels}

These states decay to two different types of channels, one is the
light meson - light baryon channel, while the other is the
$c\bar{c}$ meson - baryon channel. For the VB states, there is
another possibility to decay into PB channels, for instance,
$\bar{D}^{*}B \to \bar{D}B$. The analogous decay channels in the
$VV \to VV$ hidden charm sector driven by pseudoscalar exchange were studied in
\cite{raquel2} and found to be extremely small because of the small phase space available. Analogously, the terms involving a vector exchange contains an anomalous VVP vertex and were also found very small in \cite{raquel}. Hence, we do not
consider them here. In this subsection we only consider the decay
of these states to the light meson - light baryon channel as
depicted in the Feynman diagrams of Fig. \ref{wfe}. These diagrams
provide a negligible real part compared to the tree level
potentials. The imaginary part gives rise to a width of the
states. Hence, we only consider the effect of this box diagram on
the states found before.

\begin{figure}[htbp] \vspace{-0.cm}
\begin{center}
\includegraphics[width=0.4\columnwidth]{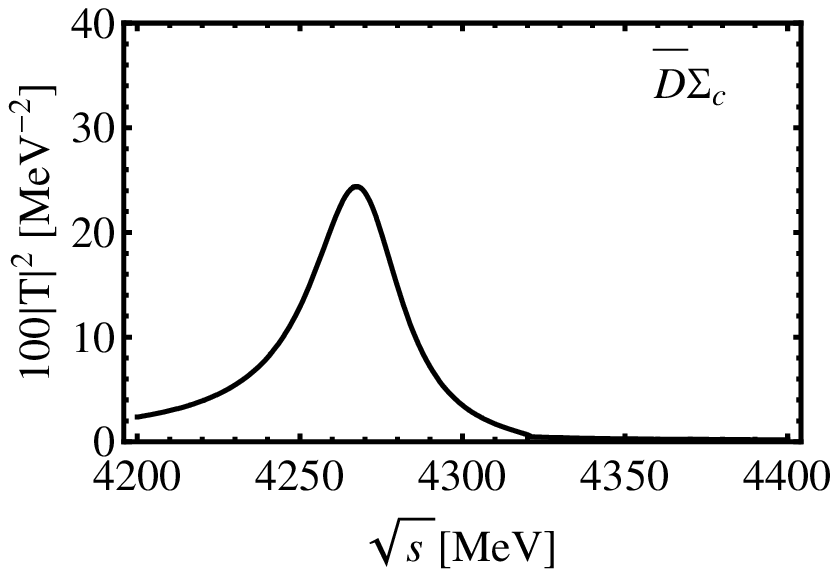}
\includegraphics[width=0.4\columnwidth]{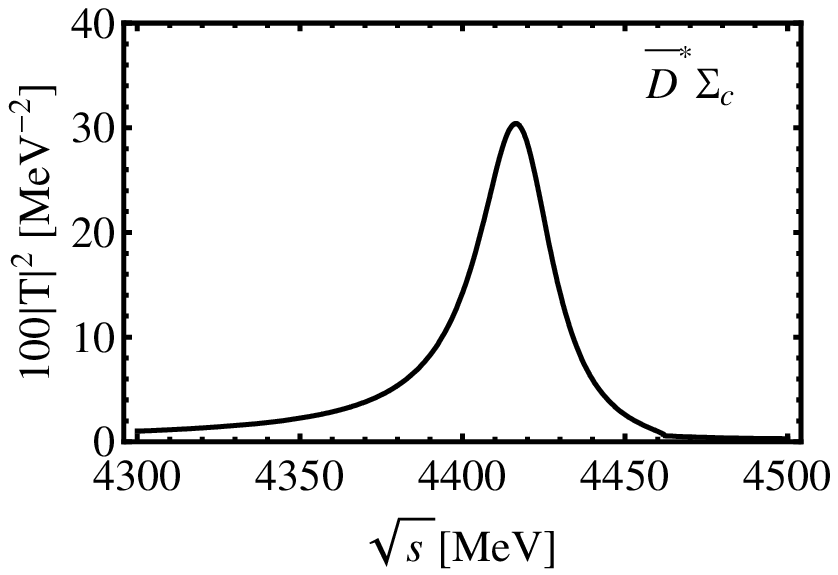}
\caption{$|T_{ii}|^2$ for the different channels in the $(I=1/2, S=0)$ sector including the box diagrams.} \label{i12s0}
\end{center}
\begin{center}
\includegraphics[width=0.4\columnwidth]{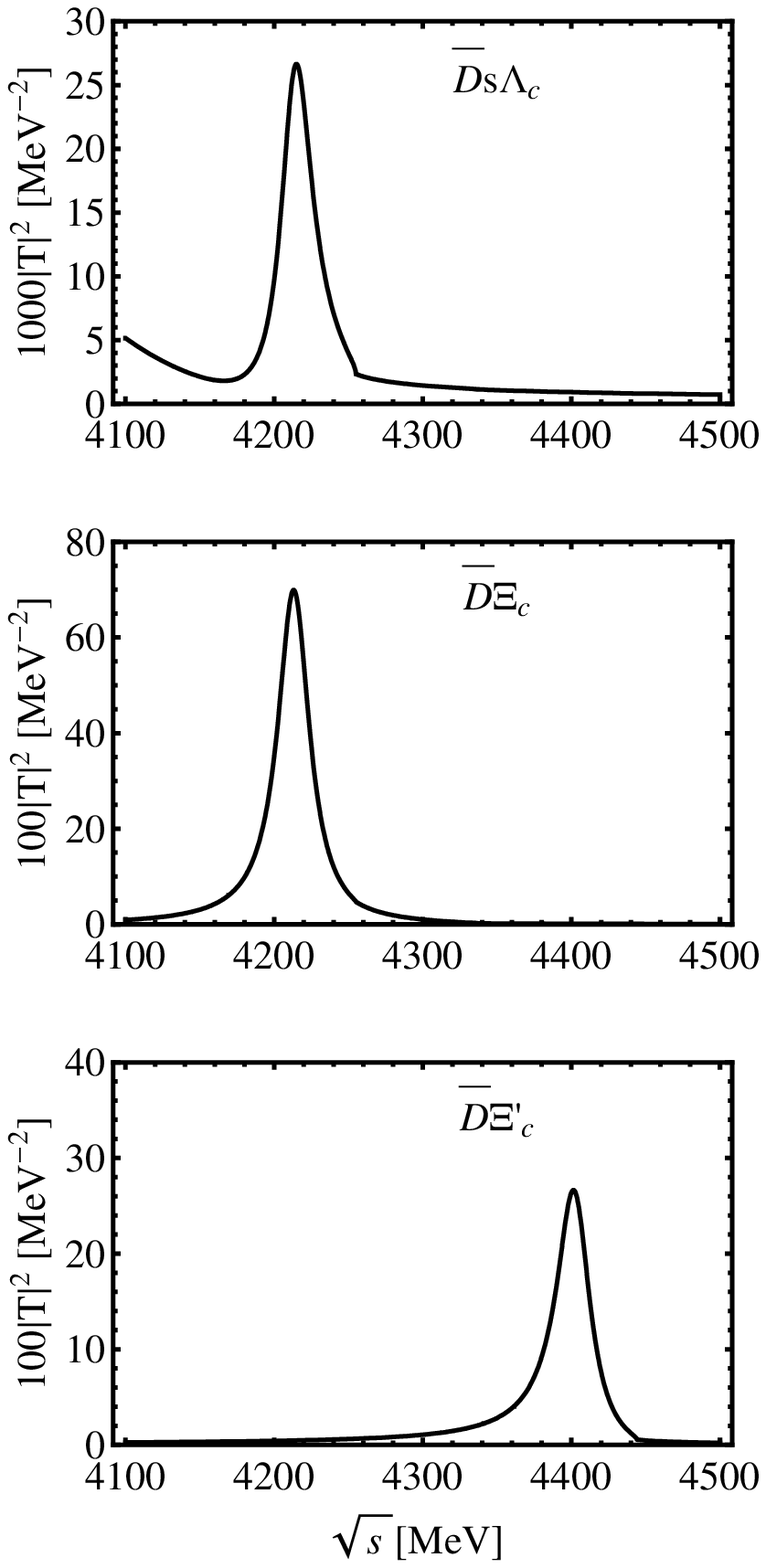}
\includegraphics[width=0.4\columnwidth]{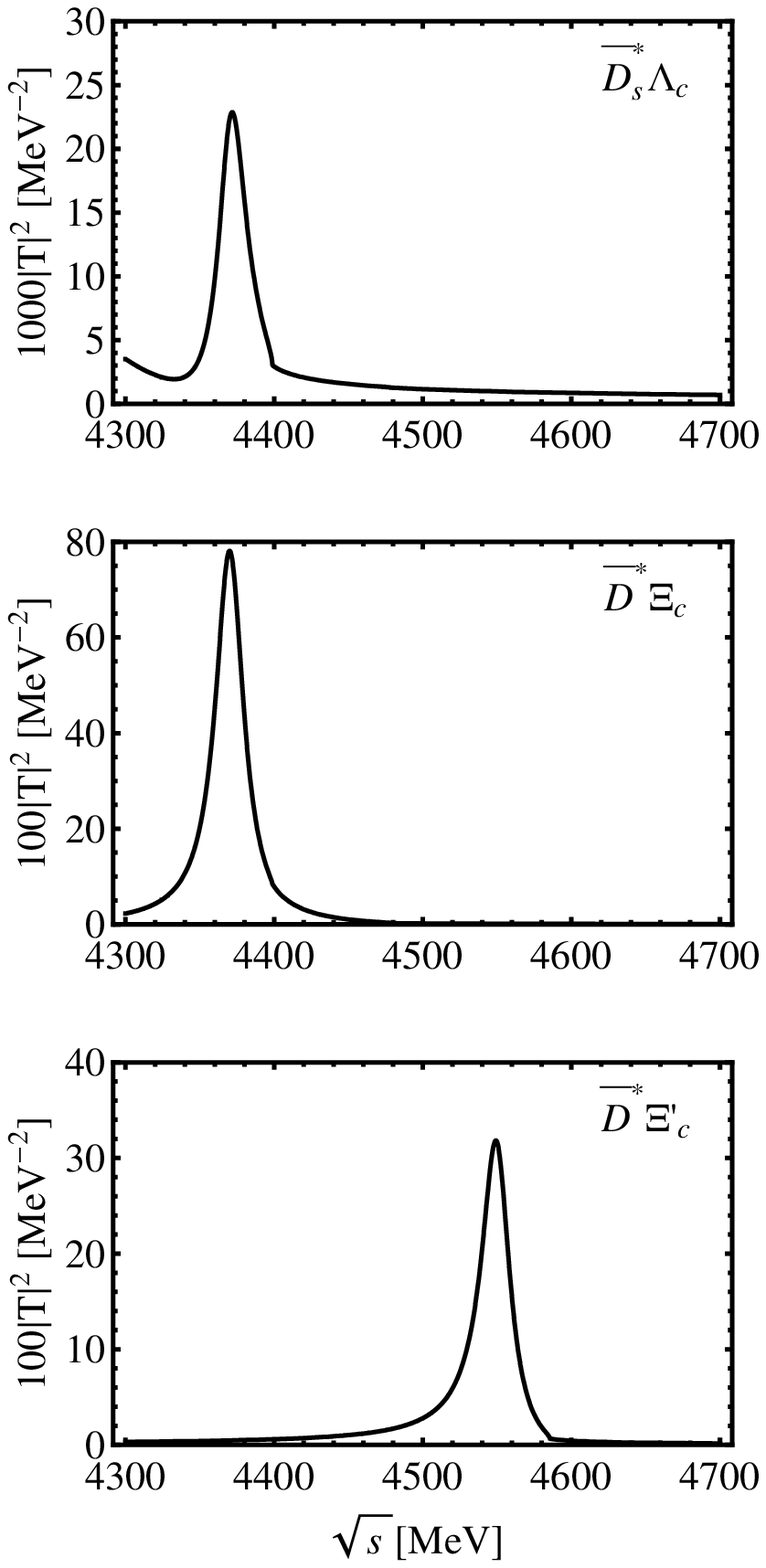}
\caption{$|T_{ii}|^2$ for the different channels in the $(I=0, S=-1)$ sector including the box diagrams.} \label{i0s1}
\end{center}
\end{figure}

In Figs. \ref{i12s0} and \ref{i0s1}, we show the results of
$|T_{ii}|^2$ as a function of $\sqrt{s}$ for the different
channels, and we list their decay widths to the different channels
for all the sectors in Tables \ref{pbwidth} and \ref{vbwidth}.
From these pictures and tables, we find that the six states are
all above $4200$ MeV. However, their widths are quite small. In
principle, one might think that the width of these massive objects
should be large because there are many channels open and there is
much phase space for decay. However, it is difficult for the
$c\bar{c}$ components to decay to the $u\bar{u}$, $d\bar{d}$ and
$s\bar{s}$ ones, something that within our model is tied to the necessity of the
exchange of a heavy - vector meson. Note that the pole positions are
obtained without including the box diagrams by extrapolating to
the complex plane. The inclusion of the box diagram renders this
extrapolation more difficult, and thus we obtain the width of the
states by plotting $|T|^2$ versus the energy with $T$ obtained in the real axis
including the box diagrams. The individual partial decay widths
are obtained including one by one the different box diagrams.

\begin{table}[ht]
      \renewcommand{\arraystretch}{1.1}
     \setlength{\tabcolsep}{0.4cm}
\begin{center}
\begin{tabular}{ccccccccc}\hline
$(I, S)$      &  $z_R$     & \multicolumn{2}{c}{Real axis} & \multicolumn{5}{c}{$\Gamma_i$ }\\
          &   & $M$ & $\Gamma$                  & \multicolumn{5}{c}{}\\
\hline
$(1/2, 0)$    &           &      &             & $\pi N$ & $\eta N$ & $\eta' N$ & $K \Sigma$ \\
          & $4269$      & $4267$ & $34.3$        & $3.8$     & $8.1 $     & $3.9$       & $17.0$\\
\hline
$(0, -1)$     &           &      &             & $\bar{K} N$      & $\pi \Sigma$  & $\eta \Lambda$ & $\eta' \Lambda$ & $K \Xi$\\
          & $4213$      & $4213$ & $26.4$        & $15.8$             & $2.9$           & $3.2 $           & $1.7$             & $2.4$     \\
          & $4403 $     & $4402$ & $28.2$        & $0 $               & $10.6$          & $7.1 $           & $3.3 $            & $5.8 $     \\
\hline\end{tabular} \caption{Pole position ($z_R$), mass ($M$),
total width ($\Gamma$), and the decay width for each particular
light meson - light baryon channel ($\Gamma_i$) for the states
from $PB\rightarrow PB$. The units are in MeV.}
 \label{pbwidth}
\end{center}
       \renewcommand{\arraystretch}{1.1}
     \setlength{\tabcolsep}{0.4cm}
\begin{center}
\begin{tabular}{ccccccccc}\hline
$(I, S)$&  $z_R$   & \multicolumn{2}{c}{Real axis} & \multicolumn{5}{c}{$\Gamma_i$  }\\
    &   & $M$ & $\Gamma$                  & \multicolumn{5}{c}{}\\
\hline
$(1/2, 0)$    &      &      &        & $\rho N$ & $\omega N$ & $K^{*} \Sigma$ \\
          & $4418$ & $4416$ & $28.4$   & $3.2$      & $10.4  $      &  $13.7$           \\
\hline
$(0, -1)$     &      &      &        & $\bar{K}^{*} N$      & $\rho \Sigma$  & $\omega \Lambda$ & $\phi \Lambda$  & $K^{*} \Xi$\\
          & $4370$ & $4371$ & $23.3 $  & $13.9 $                & $3.1  $          & $0.3 $             & $4.0$             & $1.8  $     \\
          & $4550$ & $4549$ & $23.7$   & $0 $                   & $8.8 $           & $9.1  $            & $0 $              & $5.0 $      \\
\hline
\end{tabular}\caption{Pole position ($z_R$), mass ($M$), total width ($\Gamma$), and the
decay width for each particular light meson - light baryon channel
($\Gamma_i$) for the states from $PB\rightarrow PB$. The units are
in MeV.}
 \label{vbwidth}
\end{center}
\end{table}

\subsection{Decay width to $c\bar{c}$ meson - light
baryon channels}

In this subsection we discuss the decay width of these states to
$c\bar{c}$ meson and light - baryon channels. The three states
from the $VB$ system decay to $J/\psi N$. The decay of these $VB$
states to $\eta_c N$ is also possible by means of a BBP vertex
(exchange of a pseudoscalar meson) but as we will see in the
Subsection IV. B this vertex is very small. We could also have
this decay exchanging a vector meson instead of a pseudoscalar
one, but then the amplitude would contain an anomalous VVP vertex,
which is also very small \cite{raquel}. Similarly, the decay width
of the $PB$ states to the $VB$ channels must be very small because
of the same reasons. We will consider their decay to $J/\psi N$ in
Section IV. B and we anticipate that this decay width is very
small. For these reasons, we only consider the $J/\psi N$, $J/\psi
\Lambda$  channels for the $VB$ states, and $\eta_{c}N$,
$\eta_c\Lambda$ channels in the case of states from the $PB$
system. Thus, these new channels are added to the previous
calculation in the Subsections III. A and B.

The pole positions of these states only change a bit compared to
those given in the Subsection III. A, since the potentials from
these channels are much smaller. Nevertheless, these channels
provide some extra width because, in spite of the smaller phase
space for the decay, the three momentum transfer in the propagator
of the $D^{*}(D^{*}_{s})$ exchange is much smaller than in the case
of transition to light meson - light baryon channels. The transition
potential becomes:
\begin{eqnarray}
V_{ab(PB\rightarrow \eta_{c} B)}&=&-\frac{C_{ab}
g^2}{p^{2}_{D^{*}}-M^{2}_{D^{*}}}(E_{P}+E_{\eta_{c}})\ ,
\label{pbetacb}
\end{eqnarray}
where
\begin{eqnarray}
p^{2}_{D^{*}}=M^{2}_{\eta_{c}}+M^{2}_{P}-2E_{\eta_{c}}E_{P}\ ,
\end{eqnarray}
and similarly for the $VB$ system but changing $p_{D^*}$,
$M_{D^*}$, $E_P$ and $E_{\eta_{c}}$ by $p_{D}$, $M_{D}$, $E_V$ and
$E_{J/\psi}$ respectively. Here we also neglect the three-momentum
of the final and initial particles because we consider energies
close to the threshold. We list the results in Tables
\ref{etaccoupling}, \ref{etacwidth}, \ref{jpsicoupling} and
\ref{jpsiwidth}. We observe that the coupling constants change a
bit, but what is more relevant is that these new channels give an
extra contribution to the width, smaller, but of the same order as
the one obtained previously. The relatively large decay width to
the $\eta_c N$ channel is a good feature with respect to the
possible observation of these resonances since there will be less
background in $\eta_c N$ than in $\pi N$, $\eta N$, $K\Sigma$, the
observation of the resonance in the $\eta_c N$ channel could be
favoured.

In Tables \ref{etaccoupling} and \ref{etacwidth}, the pole
positions are obtained without the box diagrams, but including the
$\eta_c N$, $\eta_c \Lambda$ channels. Now the pole positions
becomes complex because the new channels are open. We can see that
the partial decay width into these channels is approximately twice
the imaginary part of the pole position. The total widths are
again obtained by looking at the width of $|T|^2$ in the real axis
when the box diagrams are included. We would like to mention that
in the approach of \cite{Hofmann:2005sw}, which has been corrected
in \cite{Mizutani:2006vq,Tolos:2007vh}, some hidden charm states
are also found, bound by about $1000$ MeV. It is not easy to
understand such a large binding on physical grounds, which is not
supported in any case by the strength of the potentials.
\begin{table}[ht]
      \renewcommand{\arraystretch}{1.1}
     \setlength{\tabcolsep}{0.4cm}
\begin{center}
\begin{tabular}{cccccc}\hline
$(I, S)$&  $z_R$ (MeV)                   & \multicolumn{4}{c}{$g_a$}\\
\hline
$(1/2, 0)$    &             & $\bar{D} \Sigma_{c}$ & $\bar{D} \Lambda^{+}_{c}$&  $\eta_{c}N$ \\
          & $4265-11.6i$  & $2.96-0.21i$           &$-0.08+0.06i$               &  $-0.94+0.03i$ \\
          &               &  $ 2.97$               &$0.10  $                    &  $0.94 $\\
\hline
$(0, -1)$  &                & $\bar{D}_{s} \Lambda^{+}_{c}$   & $\bar{D} \Xi_{c}$ & $\bar{D} \Xi'_{c}$ & $\eta_{c}\Lambda$\\
       & $  4210-2.9i $   & $1.42-0.03i   $                   & $3.28-0.002i $      & $-0.15+0.13i  $      & $0.57+0.04i  $      \\
       &                & $1.42    $                      & $3.28  $          &$ 0.19  $           & $0.57$            \\
       &  $ 4398-8.0i$    & $0.01+0.004i $                    & $0.06-0.02i$        & $2.75-0.15i$         & $-0.73-0.07i$                   \\
       &                & $0.01  $                        & $0.06  $          & $2.75$             & $0.74 $                  \\
\hline\end{tabular} \caption{Pole position, $z_R$ and coupling
constants, $g_a$, to various channels for the states from
$PB\rightarrow PB$ including the $\eta_{c}N$ and $\eta_{c}\Lambda$
channel.}
 \label{etaccoupling}
\end{center}
   \renewcommand{\arraystretch}{1.1}
     \setlength{\tabcolsep}{0.4cm}
\begin{center}
\begin{tabular}{ccccc}\hline
$(I, S)$      &  $z_R$ (MeV)      & \multicolumn{2}{c}{Real axis} & $\Gamma_i$ \\
          &    & $M$ & $\Gamma$                  & \\
\hline
$(1/2, 0)$    &            &      &             &  $\eta_{c}N$\\
          & $4265-11.6i$ & $4261$ & $56.9$        &  $23.4$       \\
\hline
$(0, -1)$     &            &      &             &  $\eta_{c}\Lambda$\\
          & $4210-2.9i$  &$ 4209 $& $32.4 $       &$  5.8$        \\
          & $4398-8.0i$  & $4394$ & $43.3$        &  $16.3$        \\
\hline\end{tabular} \caption{Pole position ($z_R$), mass ($M$),
total width ($\Gamma$, including the contribution from the light
meson and baryon channel) and the decay widths for the $\eta_{c}N$
and $\eta_{c}\Lambda$ channels ($\Gamma_i$). The unit are in MeV}
 \label{etacwidth}
\end{center}
\end{table}

\begin{table}[ht]
      \renewcommand{\arraystretch}{1.1}
     \setlength{\tabcolsep}{0.4cm}
\begin{center}
\begin{tabular}{cccccc}\hline
$(I, S)$&  $z_R$                   & \multicolumn{4}{c}{$g_a$}\\
\hline
$(1/2, 0) $   &             & $\bar{D}^{*} \Sigma_{c}$ & $\bar{D}^{*} \Lambda^{+}_{c}$&  $J/\psi N$ \\
          & $4415-9.5i$   & $2.83-0.19i     $          &$-0.07+0.05i   $            &  $-0.85+0.02i$ \\
          &             &$ 2.83  $                 &$0.08  $                  &  $0.85$ \\
\hline
$(0, -1) $ &                & $\bar{D}^{*}_{s} \Lambda^{+}_{c}$   & $\bar{D}^{*} \Xi_{c}$ & $\bar{D}^{*} \Xi'_{c}$ & $J/\psi \Lambda$\\
       &   $4368-2.8i  $  & $1.27-0.04i     $                     &$ 3.16-0.02i $          & $-0.10+0.13i  $          & $0.47+0.04i   $     \\
       &                & $1.27 $                             & $3.16 $               & $0.16 $                & $0.47  $          \\
       &   $4547-6.4i $   & $0.01+0.004i$                         & $0.05-0.02i$            & $2.61-0.13i $            & $-0.61-0.06i   $                \\
       &                & $0.01   $                           & $0.05 $               & $2.61$                 & $0.61 $                  \\
\hline\end{tabular} \caption{Pole position ($z_R$) and coupling
constants ($g_a$) to various channels for the states from
$PB\rightarrow PB$ including the $J/\psi N$ and $J/\psi\Lambda$
channels. }
 \label{jpsicoupling}
\end{center}
       \renewcommand{\arraystretch}{1.1}
     \setlength{\tabcolsep}{0.4cm}
\begin{center}
\begin{tabular}{ccccc}\hline
$(I, S)$      &  $z_R$      & \multicolumn{2}{c}{Real axis} & $\Gamma_i$ \\
          &    & $M$ & $\Gamma$                  & \\
\hline
$(1/2, 0)$    &            &      &             &  $J/\psi N$\\
          & $4415-9.5i$  & $4412$ & $47.3$        &  $19.2$       \\
\hline
$(0, -1)$     &            &      &             &  $J/\psi\Lambda$\\
          & $4368-2.8i$  & $4368 $& $28.0 $       &  $5.4$        \\
          & $4547-6.4i $ & $4544 $& $36.6 $       &  $13.8$        \\
\hline\end{tabular} \caption{Pole position ($z_R$), mass ($M$),
total width ($\Gamma$, including the contribution from the light
meson and baryon channel) and the decay widths for the $J/\psi N$
and $J/\psi\Lambda$ channels ($\Gamma_i$). The unit are in MeV}
 \label{jpsiwidth}
\end{center}
\end{table}

\section{Production cross section in $\bar{p}p$ collisions}

\subsection{ Estimate of the $p\bar{p}\to N^{*+}_{c\bar{c}}(4265)\bar{p}$ cross section}
\begin{figure}[htpb]
\begin{center}
\includegraphics[width=0.5\columnwidth]{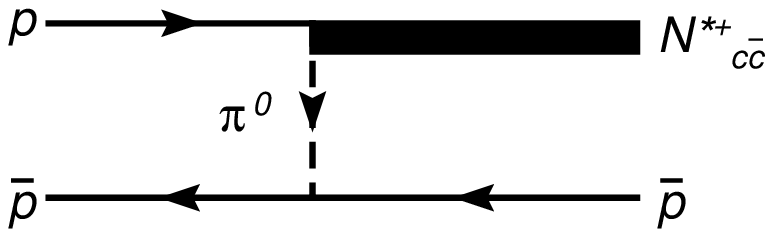}
\caption{The $p\bar{p}\to N^{*+}_{c\bar{c}} \bar{p}$ mechanism.} \label{wu0}
\end{center}
\end{figure}

We shall estimate the production cross section of these resonances
at FAIR. With a $\bar{p}$ beam of $15~GeV/c$ one has $\sqrt s= 5470
~MeV$, which allows one to observe resonances in $\bar{p} X$
production up to a mass $M_X\simeq 4538 ~MeV$. We shall make some
rough estimate of the cross section for the   $\bar{p} p \to
\bar{p} N^{*+}_{c\bar{c}}$ production for the $C=0,S=0$ resonances
that we have obtained from the pseudoscalar  baryon interaction.
Since one important decay channel of the $N^{*}_{c\bar{c}}$ is
$\pi N$, we evaluate the cross section for the mechanism depicted
in the Feynman diagram of Fig. \ref{wu0}.

The coupling of the $N^{*}_{c\bar{c}} \to \pi^0 p$ is obtained
projecting over $\pi^0 p$ the isospin state $I=1/2$, which provides
the isospin coefficient $C_I=\sqrt{1/3}$. The coupling
$N^{*}_{c\bar{c}} \to \pi N$ we get from the partial decay width
of the $N^{*}_{c\bar{c}}$ into this channel, $\Gamma_{\pi N}$
\begin{equation}
 g_{N^{*}_{c\bar{c}} \to \pi N}^2 =\frac{2\pi M_{N^{*}_{c\bar{c}}} \Gamma_{\pi N}}{M_N p_\pi^{\mathrm{on}}}\label{eq:g2}
\end{equation}
with
$p_\pi^{\mathrm{on}}=\lambda^{1/2}(M_{N^{*}_{c\bar{c}}}^2,m^2_\pi,M^2_N)/2
M_{N^{*}_{c\bar{c}}}$, the value of the on-shell pion momentum
from the $N^{*}_{c\bar{c}}\to \pi N$ decay. By taking the standard
$\pi NN$ vertex, $V_{\pi NN}= i g_{\pi} \gamma_{5} \tau^\lambda$ ($g_{\pi}\simeq 13$),
we obtain
\begin{equation}
\frac{d\sigma_{p\bar{p} \to N^{*+}_{c\bar{c}}
\bar{p}}}{d\mathrm{cos}\theta}=\frac{g^2_\pi}{4}\frac{M_X^2}{s}\frac{\Gamma_{\pi
N}C^{2}_{I}}{p_\pi^{\mathrm{on}}}\frac{2p.p'-2 M^2}{(2 M^2-\sqrt{s}
E(p') + 2\vec{p}.\vec{p}\,')^2} \frac{p'}{p} \label{etacn}
\end{equation}
where $p, p' $ are the initial, final momenta of the $\bar{p}$ in
the center of mass frame ( of the order of 2570, 620 MeV/c for
$M_X\simeq 4300$ MeV). The biggest cross section corresponds to
the forward $\bar{p}$ direction, which is the most indicated for
the search. If we are interested in searching for these
resonances, looking for $\bar{p}$ forward is the most
recommendable measurement and one should look for a bump into the
$d \sigma/d \mathrm{cos} {\theta} d M^{2}_I$ magnitude, where
$M_I$ is the invariant mass of the $\pi N$ coming from the decay
of the produced $N^{*+}_{c\bar{c}}$ state. Assuming a Lorentzian
shape for this resonance, with total width
$\Gamma_{N^{*+}_{c\bar{c}}}$, we would obtain at the peak of the
$\pi N$ distribution

\begin{equation}
 \frac{ d\sigma_{p\bar{p} \to N^{*+}_{c\bar{c}}(4265)\bar{p}\to \pi N \bar{p}} }{d\mathrm{cos}\theta dM^{2}_I}=\frac{1}{\pi}\frac{1}{M_{N^{*+}_{c\bar{c}}}\Gamma_{\mathrm{tot}}}\frac{d\sigma_{p\bar{p} \to N^{*+}_{c\bar{c}}\bar{p}}}{d \mathrm{cos}\theta}\frac{\Gamma_{\pi N}}{\Gamma_{\mathrm{tot}}}\label{eq:dgam}
\end{equation}
which leads to the following cross section: $0.13$ $\mu b$/GeV$^2$
for $N^{*+}_{c\bar{c}}(4265)$.

In the above calculation, we did not consider the form factor for
the $\pi NN$ vertex. The form factor is:
\begin{equation}
F_{pp\pi}=\frac{\Lambda^{2}_{\pi}-m^{2}_{\pi}}{\Lambda^{2}_{\pi}-p^{2}_{\pi}}.\label{eq:fnnpi}
\end{equation}
with the $\Lambda_{\pi}=1.3GeV$. We can multiply by $F^{2}_{pp\pi}$
the cross section in the Eq. (\ref{eq:dgam}) and we find about
$0.05$ $\mu b$/GeV$^2$.

Because in such high energy transfer reaction the one-pion exchange
with the monopole off-shell form factor of Eq.(\ref{eq:fnnpi}) may
not be a good approximation, here we also make a calculation with
the Reggeon exchange. Using a Reggeon propagator
$R_{\pi}(s,t)$~\cite{anisovich} instead of the usual pion
propagator. Then the Eq.(\ref{etacn}) becomes
\begin{equation}
 \frac{d\sigma_{p\bar{p} \to N^{*+}_{c\bar{c}} \bar{p}}}{d\mathrm{cos}\theta}=\frac{g^2_\pi}{4}\frac{M_X^2}{s}\frac{\Gamma_{\pi N}C^{2}_{I}}{p_\pi^{\mathrm{on}}}(\sqrt{s} E(p') - 2pp'\mathrm{cos}\theta-2
 M^2_{N})|R_{\pi}(s,t)|^{2}\frac{p'}{p},
\label{piregge}
\end{equation}
where
\begin{eqnarray}
R_{\pi}(s,t)&=&-\frac{\pi}{2}\alpha'_{pi}(t)\mathrm{exp}(-i\frac{\pi}{2}\alpha_{\pi}(t))\frac{(s/s_{0})^{\alpha_{\pi}(t)}}{\mathrm{sin}(\frac{\pi}{2}\alpha_{\pi}(t))\mathrm{\Gamma}(\frac{\alpha_{\pi}(t)}{2}+1)},\\
 \alpha_{\pi}(t)&=&-0.015+0.72t,\\
 \alpha'_{\pi}(t)&=&0.72,\\
 t&=&2 M^2_{N}-\sqrt{s} E(p') + 2pp'\mathrm{cos}\theta,
\end{eqnarray}
with the slope parameter in the units of ($GeV^{-2}$). When $t \to
m^2_{\pi}$, we can get $\alpha_{\pi}=0$ and $R_{\pi} \sim
\frac{1}{m^2_{\pi}-t}$. From Ref.\cite{anisovich}, the order of
$s_0$ is about $2-20GeV^{-2}$. To narrow down its range, we use the
information of $pp$ collision with $\sqrt{s}<3$ GeV where the
one-pion exchange can reproduce experimental data reasonably
well~\cite{ouyang}. Demanding the Reggeon propagator to give similar
results as the usual $\pi$ propagator for $\sqrt{s} < 3$ GeV, we
have $s_0 \simeq 5-10GeV^{-2}$. Then by using Reggeon propagator for
the maximum PANDA energy $\sqrt{s} < 5.47$ GeV, we get the cross
section to be about $0.006 \sim 0.017$ $\mu b$/GeV$^2$ corresponding
to $s_0 = 5\sim 10$ GeV$^{-2}$. This is about a factor $3\sim 9$
smaller than the result by one-pion exchange.

\begin{figure}[htbp] \vspace{-0.cm}
\begin{center}
\includegraphics[width=0.7\columnwidth]{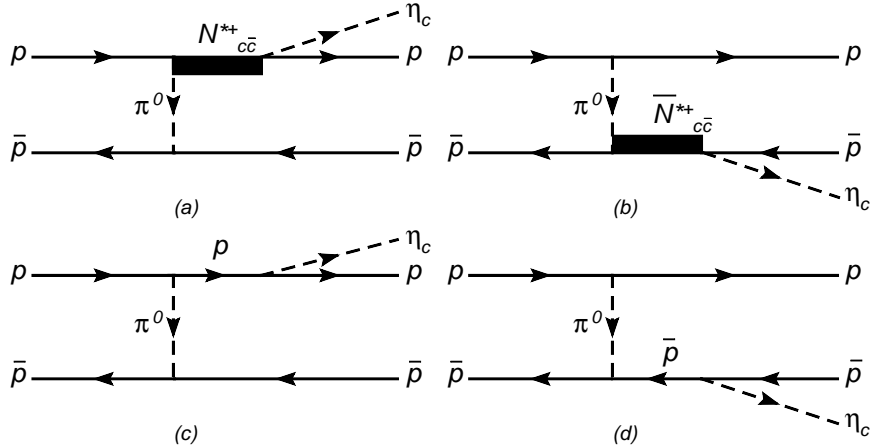}
\caption{The different Feynman diagrams of the reaction $p\bar{p}
\rightarrow p\bar{p} \eta_{c}$} \label{fe3}
\end{center}
\end{figure}

Then we can estimate the cross section of $p\bar{p} \rightarrow
p\bar{p} \eta_{c}$. The different Feynman diagrams for this reaction
are shown in the Fig.~\ref{fe3}. Using Eq. (\ref{eq:dgam}) and
$\Gamma_{\eta_c p}$ of the resonance instead of $\Gamma_{\pi N}$ we
can obtain the differential cross section at the peak of the
resonance, corresponding to the resonant mechanism of Fig. \ref{fe3}
a), and it is about $0.8$ $\mu b$/GeV$^2$ without form factor and
$0.3$ $\mu b$/GeV$^2$ with the form factor, and $0.04 \sim 0.10$
$\mu b$/GeV$^2$ with the Reggeon propagator. This magnitude is of
about the same order of magnitude as typical cross sections measured
for $d \sigma/d \mathrm{cos} {\theta} d M^{2}_I$ in the $pd \to p d
\pi^0 \pi^0$ or $p p\to d\pi^+\pi^0$ reaction
\cite{clement,clement2}. In order to see the role played by the
hidden charm resonance in this process we can compare it with the
cross section coming from a standard mechanism of Fig.
\ref{fe3}(c,d). The vertex of $pp\eta_{c}$ is used by
\begin{eqnarray}
 {\cal L}_{\eta_{c}p\bar{p}}&=&g_{\eta_{c}p\bar{p}}\bar{u}_{p}\mathbf{\gamma^\mu\gamma^{5}\partial_{\mu}\psi_{\eta_{c}}}v_{\bar{p}},
\label{pppi}
\end{eqnarray}
where $g_{\eta_{c}p\bar{p}}$ can be calculated from the reaction
$\eta_{c} \to p\bar{p}$ by
\begin{eqnarray}
 g_{\eta_{c}p\bar{p}}=\sqrt{\frac{\pi \Gamma_{\eta_{c}}Br_{\eta_{c}p\bar{p}}}{|p^{on}_{p}|m^{2}_{p}}}. \label{i}
\end{eqnarray}
where the
$p^{on}_{p}=\lambda^{1/2}(m_{\eta_{c}}^2,M^2_p,M^2_{\bar{p}})/2m_{\eta_{c}}$
the value of the on-shell $p$ momentum from the $\eta_{c} \to
p\bar{p}$ decay. And the width $\Gamma_{\eta_{c}}=26.7MeV$ and the
branch ratio $Br_{\eta_{c}p\bar{p}}=1.3\times10^{-3}$ are both from
PDG. The form factor of the vertex $NN\pi$ is also used
Eq.(\ref{eq:fnnpi}). We also add the form factors for
$N^{*}_{c\bar{c}}$ and $p$ exchange in the Fig.\ref{fe3}:
\begin{eqnarray}
F_{p}&=&\frac{\Lambda_{p}^{4}}{\Lambda_{p}^{4}+(p^{2}_{p}-m^{2}_{p})^{2}},\label{eq:fp}\\
F_{N^{*}_{c\bar{c}}}&=&\frac{\Lambda_{N}^{4}}{\Lambda_{N}^{4}+(p^{2}_{N^{*}_{c\bar{c}}}-m^{2}_{N^{*}_{c\bar{c}}})^{2}}.
\label{eq:fns}
\end{eqnarray}
Here $\Lambda_{p}=\Lambda_{N}=0.8GeV$.

\begin{figure}[htbp] \vspace{-0.cm}
\begin{center}
\includegraphics[width=0.6\columnwidth]{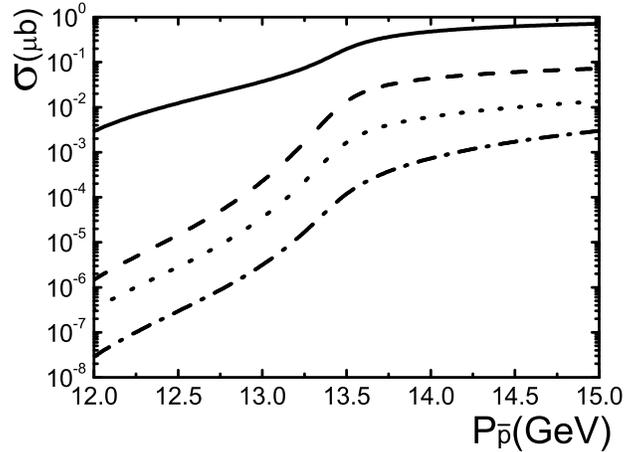}
\caption{The total cross section vs the beam momentum of $\bar{p}$
for $p\bar{p} \to p\bar{p} \eta_{c}$. The solid and dashed lines are
calculated by one-pion exchange without and with form factors,
respectively. The dot-dashed and dotted lines are results by Reggeon
propagator with $s_0=5$ and $10$ GeV$^{-2}$, respectively.}
\label{cross}
\end{center}
\end{figure}

\begin{figure}[htbp] \vspace{-0.cm}
\begin{center}
\includegraphics[width=0.38\columnwidth]{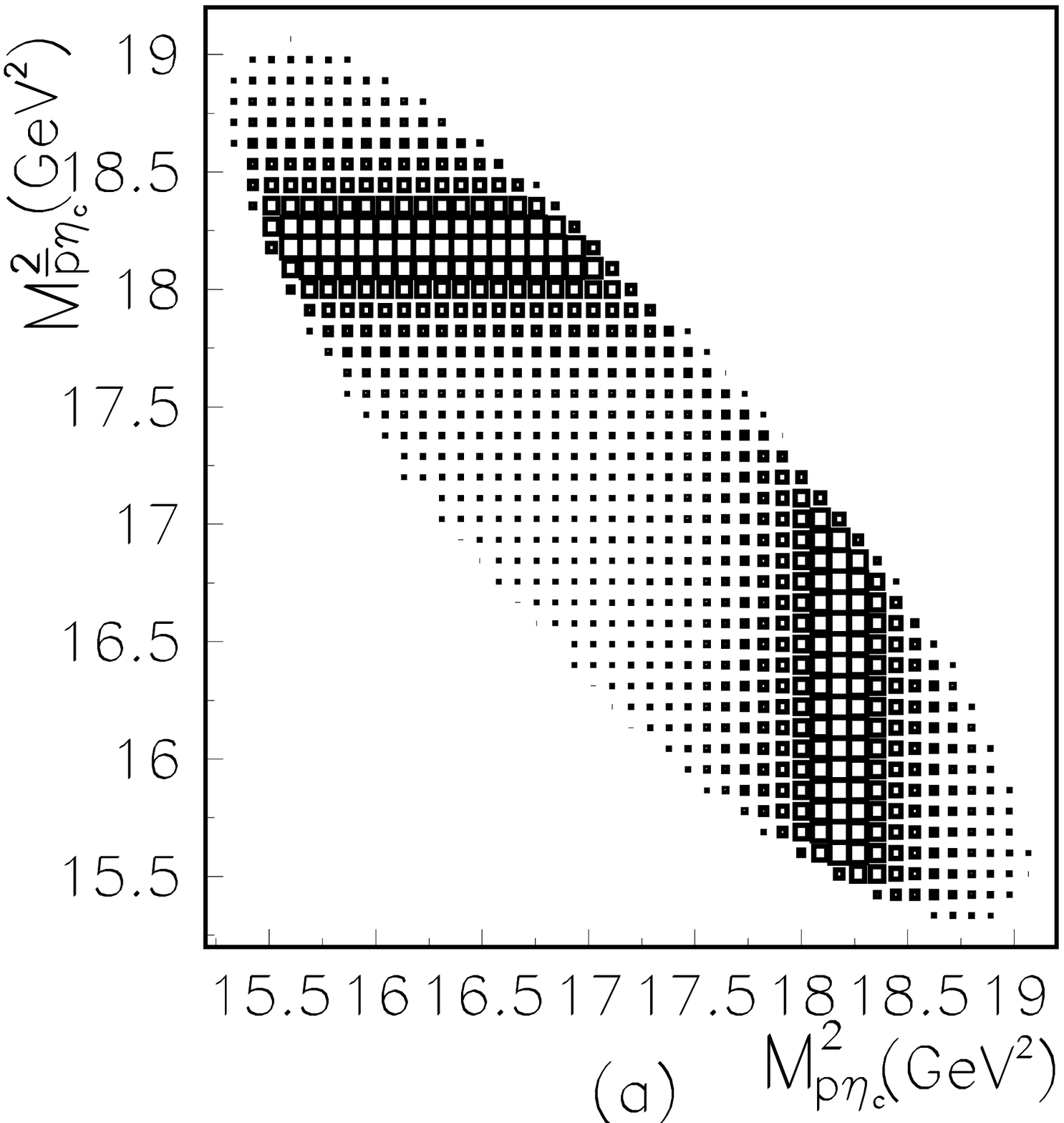}
\includegraphics[width=0.38\columnwidth]{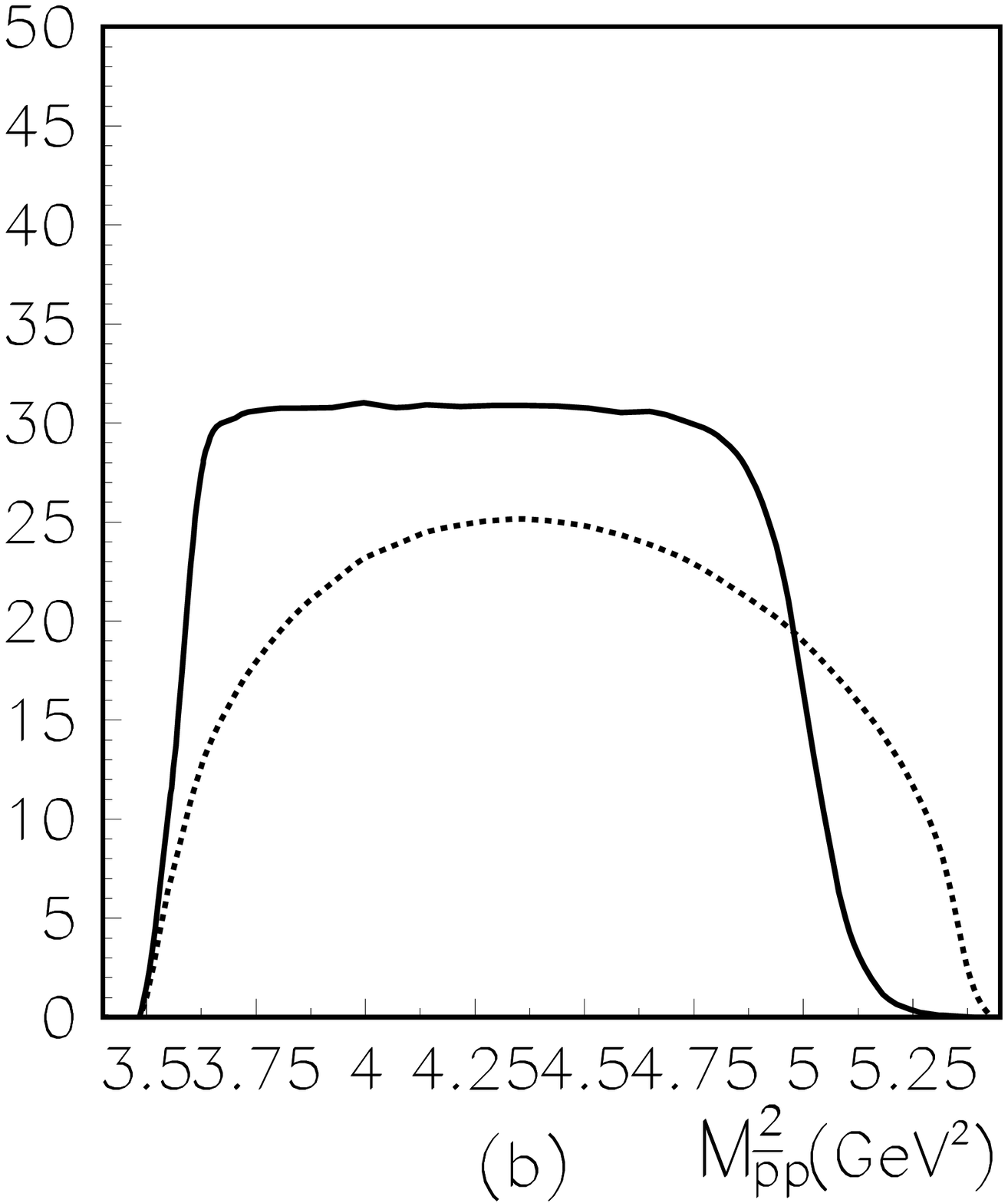}
\includegraphics[width=0.38\columnwidth]{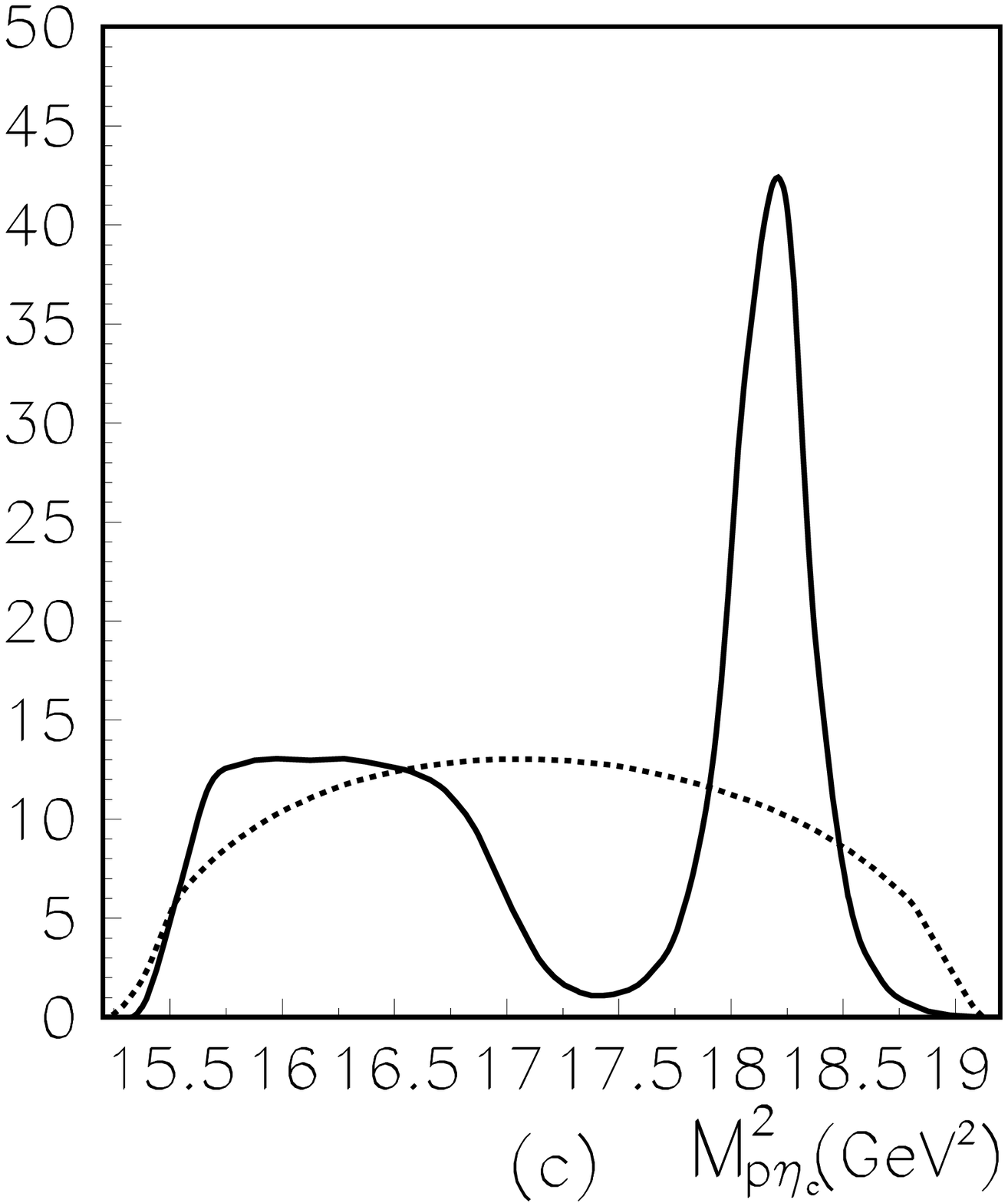}
\includegraphics[width=0.38\columnwidth]{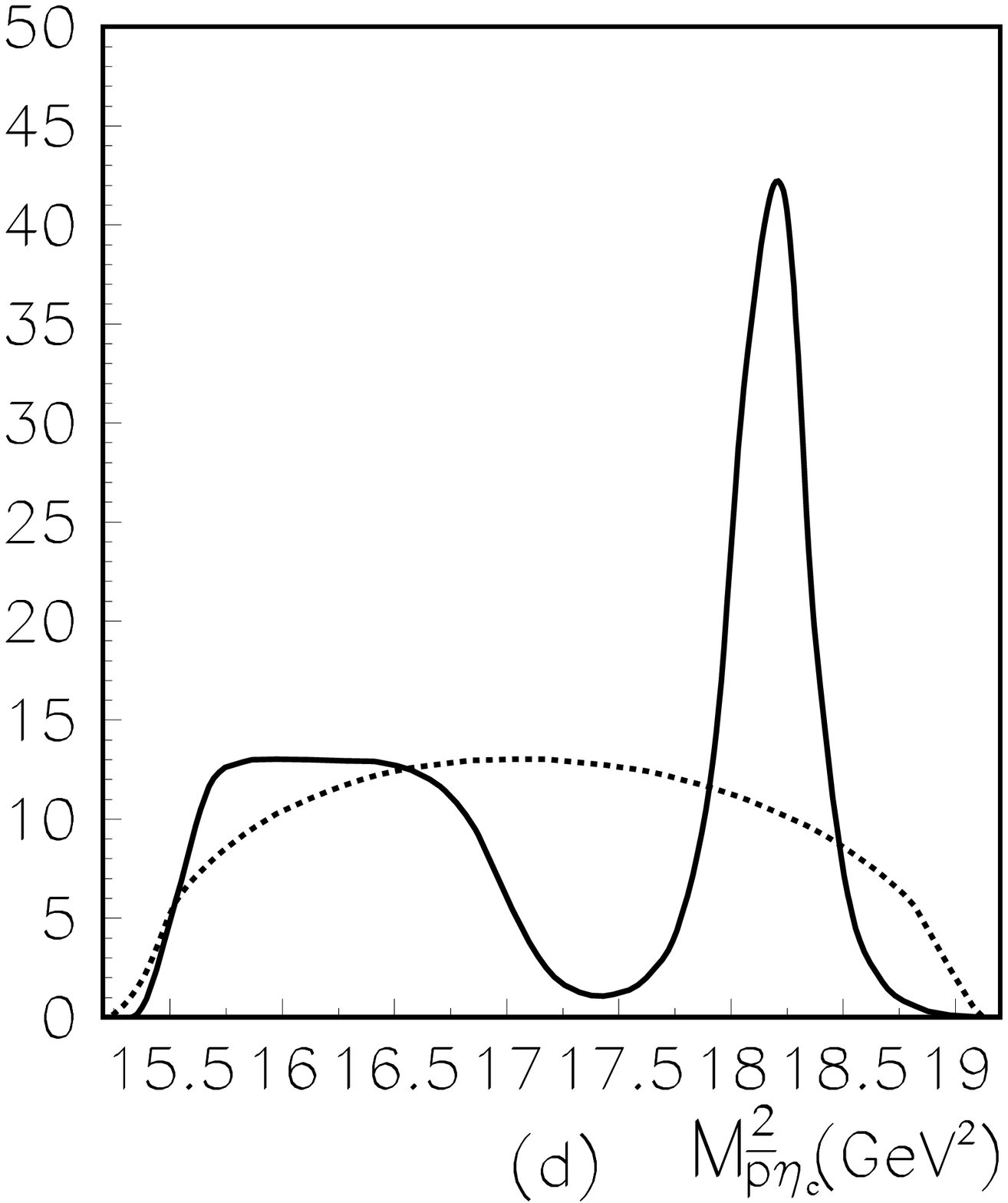}
\caption{The Dalitz plot(a), the invariant mass spectrum of
$p\bar{p}$(b), $p \eta_{c}$(c) and $\bar{p} \eta_{c}$(d) for the
reaction $p\bar{p} \rightarrow p\bar{p} \eta_{c}$ at the beam
momentum of $\bar{p}$ being 14.00GeV at lab system.} \label{fetac}
\end{center}
\end{figure}

Through the calculation, the contributions from Fig. \ref{fe3} (c),
(d) are very small, almost $10^{-4}\mu b$, the main contribution
comes from the $N^{*}_{c\bar{c}}$. The total cross section is about
$0.0029\mu b$, $0.013\mu b$, $0.072\mu b$ and $0.71\mu b$ for a
$\bar p$ beam of 15 GeV/c as shown in Fig.\ref{cross}, corresponding
to the Reggeon propagator with $s_0=5$ GeV$^{-2}$ and $s_0=10$
GeV$^{-2}$, the usual $\pi$ propagator with and without form
factors. Note that the integrated cross section involves finite
angles, rather than zero in the forward direction considered before,
where the effect of the form factor is more important. The Dalitz
plot, the invariant mass spectrum of $p \eta_{c}$, $\bar{p}
\eta_{c}$ and $p\bar{p}$ are all shown in Fig.~\ref{fetac} where the
peaks of $N^{*}(4269)$ are very clear.

\subsection{$J/\psi$  production in $\bar{p} p \to \bar{p} p J/\psi$.}
\begin{figure}[htpb]
\begin{center}
\includegraphics[width=0.7\columnwidth]{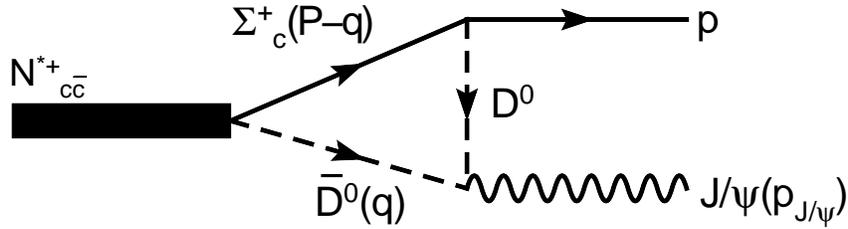}
\caption{$p J/\psi$ going to the resonance
$N^{*+}_{c\bar{c}}(4265)$. } \label{wu3}
\end{center}
\end{figure}

Another estimate that we want to do is the cross section for
$J/\psi$ production in the $\bar{p} p \to \bar{p} p J/\psi$ reaction
around the energy of the $N^*(4265)$ excitation. We use again Eq.
(\ref{eq:dgam}) but we need to evaluate $\Gamma_{J/\psi p}$. This
requires a different formalism to the one used so far. The mechanism
for $R\to J/\psi p$ is obtained by analogy to the work done in
\cite{raquel,geng} where the transition from vector - vector to
pseudoscalar - pseudoscalar states is done. Concretely, given the
fact that the $N^{*+}_{c\bar{c}}(4265)$ is basically a $\bar{D}
\Sigma_c$ molecule in our approach, we obtain the coupling of the
resonance $N^{*+}_{c\bar{c}}(4265)$ to $J/\psi p$ through the
diagram of Fig. \ref{wu3}.

This diagram requires the coupling of $N^{*+}_{c\bar{c}}(2465)$ to
the $\bar{D} \Sigma_c$ state in $I=1/2$, and the transition
$J/\psi p\to \bar{D}\Sigma_c$ which is mediated by the $\bar{D}$
meson that comes from the coupling of $J/\psi$ to $D\bar{D}$. The
diagram also involves the $DN\Sigma_c$ coupling which has been
studied in \cite{raquelmedio}.

The $J/\psi \to D\bar{D}$ coupling can be obtained from the Lagrangian
\begin{equation}
{\cal L}_{PPV}=-ig\langle V^\mu[P,\partial_\mu P]\rangle \ ,
\end{equation}
used in Section II, with $g=M_V/2 f$ and $f=93$ MeV, which leads
to
\begin{equation}
-it_{J/\psi D\bar{D}}=i\,2 g \, q_\mu \epsilon^\mu\ .
\end{equation}
The vertex $DN\Sigma_c$ is obtained from \cite{raquelmedio} and
has the form
\begin{equation}
-iV_{D^0 p\Sigma^+_c}=\vec{\sigma}\cdot \vec{q}\,'(1-\frac{q'^0}{2 M'})\beta \frac{D-F}{2f}
\end{equation}
with $\beta=1$ and $q'^0$, $\vec{q}\,'$, the incoming energy,
momentum of the $D$ meson and $M'$ the mass of the $\Sigma_c$. For
$D$ and $F$ we take the standard values $D=0.8$ and $F=0.46$
\cite{Borasoy,Close,Borasoy2}. Thus,
\begin{equation}
-it_{D^0p\Sigma^+_c}=\frac{0.26}{2f}\vec{\sigma}\cdot \vec{q}\,'
\end{equation}

We need the $I=1/2$ state of $\bar{D}\Sigma_c$ given with our
phase convention by
\begin{equation}
\vert \bar{D}\Sigma_c;1/2,1/2\rangle=\sqrt{\frac{2}{3}}D^-\Sigma_c^{++}+\frac{1}{\sqrt{3}}\bar{D}^0\Sigma^+_c\ .
\end{equation}
The other possible vertex, the $D^+p\Sigma^{++}_c$ vertex, is
$\sqrt{2}$ times the $D^0p\Sigma^+_c$ one. With all these
ingredients one obtains
\begin{eqnarray}
t_{J/\psi p\to R}&=&2\sqrt{3} \,g \int \frac{d^4 q}{(2\pi)^4}
\frac{0.26}{2 f}\vec{\epsilon}\cdot \vec{q}\,
\vec{\sigma}\cdot\vec{q}\,\frac{M_{\Sigma_c}}{E_{\Sigma_c}(q)}\,\frac{1}{q^2-m^2_D+i\epsilon}\nonumber\\&\times&\frac{1}{(p_J-q)^2-m^2_D+i\epsilon}\frac{1}{P^0-q^0-E_{\Sigma_c}(q)+i\epsilon}F(q)\
,\label{eq:dps}
\end{eqnarray}
where we use a form factor
$F(q)=\frac{\Lambda^2}{\Lambda^2+\vec{q}\,^2}$ with $\Lambda=1.05$
GeV \cite{raquelmedio} in the integral of Eq. (\ref{eq:dps}). Upon
neglecting the small three momenta $\vec{p}_{J/\psi}$ compared to
the $J/\psi$ mass and performing the $q^0$ integral, Eq.
(\ref{eq:dps}) can be written as
\begin{eqnarray}
-it_{J/\psi p\to R}&=&-\frac{1}{\sqrt{3}}  \frac{0.26}{
f}\,g\,\vec{\sigma}\cdot \vec{\epsilon}\,\int \frac{d^3
q}{(2\pi)^3}\vec{q}\,^2\,\frac{M_{\Sigma_c}}{E_{\Sigma_c}(q)}\,
\frac{1}{2\omega_D(q)}\,\frac{1}{p^0_J+2\omega_D(q)}\,\frac{1}{p^0_J-2\omega_D(q)}\,\nonumber\\&\times&\frac{1}{P^0-p^0_J-\omega_D(q)-E_{\Sigma_c}(q)}\,\frac{1}{P^0-\omega_D(q)-E_{
\Sigma_c}(q)+i\epsilon}\,\nonumber\\&\times&\lbrace2
(P^0-\omega_D(q)-E_{\Sigma_c}(q)-p^0_J-2\omega_D(q)\rbrace\
,\label{eq:t}
\end{eqnarray}
where $\omega_D(q)=\sqrt{q^2+m_D^2}$ and
$E_{\Sigma_c}(q)=\sqrt{q^2+m^2_{\Sigma_c}}$. The width of
$N^{*+}_{c\bar{c}}\to J/\psi p$ is now given by
\begin{equation}
\Gamma=\frac{1}{2\pi} \frac{M_p}{M_R} p \vert\tilde{t}_{J/\psi p\to R}\vert^2
\end{equation}
where $\tilde{t}_{J/\psi p\to R}$ means $t_{J/\psi p\to R}$
omitting the $\vec{\sigma}\cdot \vec{\epsilon}$ operator. We take
$P^0=M_R=4265$ MeV and
$p=\lambda^{1/2}(M^2_R,M^2_{J/\psi},M_p^2)/2M_R$, while $M_p$
stands for the mass of the proton. By using the form factor of
\cite{raquelmedio}, we get
\begin{equation}
\Gamma_{R\to J/\psi p}=0.01\,\, \mathrm{MeV}\ ,
\end{equation}
with admitted uncertainties of the order of a factor two. Since
$\Gamma_{\pi N}$ of the $N^{*+}_{c\bar{c}}(4265)$ was of the order
of $3.8$ MeV, now the cross section is about a factor $400$
smaller than before. Yet, the fact that the background for $J/\psi
p$ production is also smaller might compensate for it. But, from
what we have said before, the cross section for $\eta_c p$
production is much bigger.

On the other hand, for the resonances made out by $VB$, the
$J/\psi p$ production cross sections are larger. One can repeat
the calculations in this case. We sketch the derivation below.

\begin{figure}[htpb]
\begin{center}
\includegraphics[width=0.5\columnwidth]{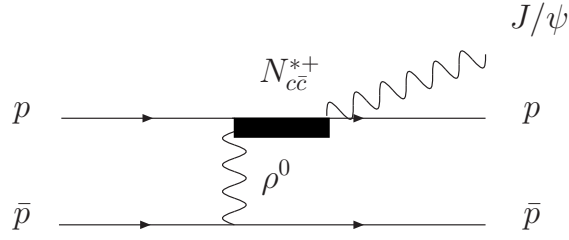}
\caption{The $p\bar{p}\to J/\psi p\bar{p}$ mechanism throughout the resonance $N^{*+}_{c\bar{c}}$ } \label{wu1}
\end{center}
\end{figure}

\begin{figure}[htbp] \vspace{-0.cm}
\begin{center}
\includegraphics[width=0.55\columnwidth]{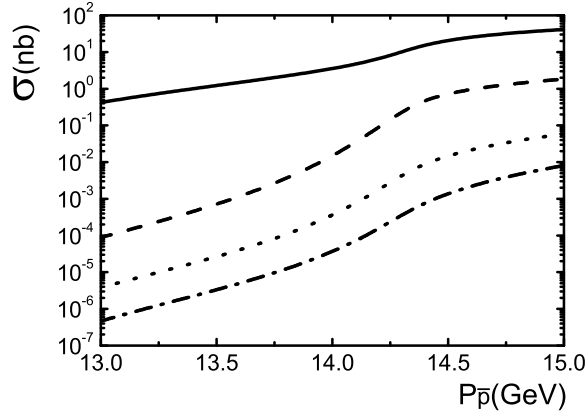}
\caption{The total cross section vs the $\bar{p}$ beam momentum for
$p\bar{p} \to p\bar{p} J/\psi$. The solid and dashed lines are
calculated by $\rho$-meson exchange without and with form factors,
respectively. The dot-dashed and dotted lines are calculated by
Reggeon propagator with $s_0=5$ and 8 GeV$^{-2}$, respectively.}
\label{JpsiN}
\end{center}
\end{figure}

We shall make the estimate based upon the mechanism of
the Feynman diagram of Fig. \ref{wu1},
and we will consider the resonance $N_{c\bar{c}}^*(4418)$ coming
from the interaction of vector mesons with baryons, one of which
channels is $J/\psi p$, which was considered in the Subsection III.
C as seen in Table \ref{jpsicoupling}. By adding this new channel we
found $g_{X J/\psi N}= 0.85$. Assuming the dominant decay channels
of $N^*$ as $\rho N$(For $\rho^{0} N$, it should be added
$C_{I}=1/\sqrt{3}$) and dominance of the $\gamma^0$ term in the
$\rho^0 p \bar{p}$ vertex, which goes then as $g \gamma^0 / \sqrt
2$, and $g=M_{v}/2f$, we obtain now
\begin{equation}
\frac{d\sigma_{p\bar{p} \to N^{*}_{c\bar{c}}(4418)
\bar{p}}}{d\mathrm{cos}\theta}=\frac{g^2}{4}\frac{M^2_X}{s}\frac{\Gamma_{\rho
N}C^{2}_{I}}{p^{\mathrm{on}}_\rho}\frac{E(p')E(p)+\vec{p}.\vec{p}\,'+M^2}{(2
M^2-\sqrt{s}
E(p')+2\vec{p}.\vec{p}\,'-M^2_\rho)^2}\frac{p'}{p}\label{N4418}
\end{equation}
with  $p',p$ the $\bar{p}$ outgoing, incoming momenta in the center
of mass frame, and $p_{\rho}^{\mathrm{on}}$ the $\rho$ momentum in
the $N^*_{c\bar{c}}(4418)$ decay into $\rho N$. By means of Eq.
(\ref{N4418}) and the width of $N^*_{c\bar{c}}(4418)\to J/\psi p$,
we can calculate the cross section of the reaction $p\bar{p} \to
J/\psi p\bar{p}$ multiplying the cross section of Eq. (\ref{N4418})
by the branching ratio of the resonance for the decay into $J/\psi
p$. As one can see in Fig. \ref{JpsiN}, this cross section is of the
order of $2\sim 37$ $n b$ for a $\bar p$ beam of 15 GeV/c, depending
on whether one includes or not the form factors. And for the dashed
line, we also give the form factor for the $NN\rho$ vertex and
$N^*_{c\bar{c}}(4418)$ as follows:
\begin{eqnarray}
F_{pp\pi}&=&\frac{\Lambda^{2}_{\rho}-m^{2}_{\rho}}{\Lambda^{2}_{\rho}-p^{2}_{\rho}}.\\
F_{N^{*}_{c\bar{c}}}&=&\frac{\Lambda_{N}^{4}}{\Lambda_{N}^{4}+(p^{2}_{N^{*}(4418)}-m^{2}_{N^{*}(4418)})^{2}}.
\end{eqnarray}
with $\Lambda_{\rho}=1.3GeV$ and $\Lambda_{N}=0.8GeV$.

\begin{figure}[htpb]
\begin{center}
\includegraphics[width=0.5\columnwidth]{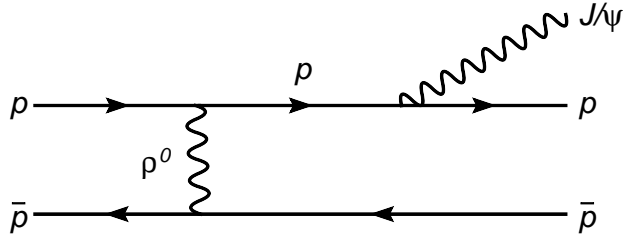}
\caption{The standard $p\bar{p}\to J/\psi p\bar{p}$ mechanism. }
\label{wu2}
\end{center}
\end{figure}

This cross section is larger than the one we would obtain from the
standard mechanism of Fig. \ref{wu2}, which can be evaluated in
analogy to the case of Fig. \ref{fe3}. Once again, using Eq.
(\ref{eq:dgam}) and $\Gamma_{J/\psi p}$ of the resonance instead of
$\Gamma_{\pi N}$ we can obtain the differential cross section of the
peak of the resonance: $6\sim 50$ nb/GeV$^{2}$.

For the same reasons as for the $N^{*}_{c\bar{c}}(4265)$ production,
we also consider the Reggeon exchange here. The Eq.(\ref{N4418})
becomes as follows:
\begin{equation}
\frac{d\sigma_{p\bar{p} \to N^{*}_{c\bar{c}}(4418)
\bar{p}}}{d\mathrm{cos}\theta}=\frac{g^2}{4}\frac{M^2_X}{s}\frac{\Gamma_{\rho
N}C^{2}_{I}}{p^{\mathrm{on}}_\rho}(E(p')E(p)+pp\,'\mathrm{cos}\theta+M^2_{N})|R_{\rho}(s,t)|^2\frac{p'}{p}
\label{piregge1}
\end{equation}
where
\begin{eqnarray}
R_{\rho}(s,t)&=&-\frac{\pi}{2}\alpha'_{\rho}(t)\mathrm{exp}(-i\frac{\pi}{2}\alpha_{\rho}(t))\frac{(s/s_{0})^{\alpha_{\rho}(t)-1}}{\mathrm{cos}(\frac{\pi}{2}\alpha_{\rho}(t))\mathrm{\Gamma}(\frac{\alpha_{\rho}(t)}{2}+\frac{1}{2})}\\
 \alpha_{\rho}(t)&=&0.5+0.83t\\
 \alpha'_{\rho}(t)&=&0.83\\
 t&=&2 M^2_{N}-\sqrt{s} E(p') + 2pp'\mathrm{cos}\theta
\end{eqnarray}
When $t \to m^2_{\rho}$, $\alpha_{\rho}=1$ and $R_{\rho} \sim
\frac{1}{m^2_{\rho}-t}$. For the estimation the value of $s_0$ is
taken to be from 5 GeV$^{-2}$ to 8 GeV$^{-2}$, such that
$|R_{\rho}(s,t)|^2$ gives almost the same results as
$\frac{1}{(m^2_{\rho}-t)^2}$ when $\sqrt{s} < 3$GeV. By using
Reggeon propagator, the total cross section is about $0.008\sim
0.06$ nb for a $\bar p$ beam of 15 GeV/c. This is about two orders
of magnitude smaller than the result by $\rho$-meson exchange.

From the calculation above, we find that the cross section of this
reaction is about two orders of magnitude smaller than that of the
reaction $p\bar{p} \rightarrow p\bar{p} \eta_{c}$, but it could be
also appropriate to find $N^{*}(4418)$ because the $J/\psi$ has a
large branching ratio to decay into lepton channels which are much
easier to detect than hadron channels.

Finally let us discuss the possibility of measurement of this
reaction in the experiments. The PANDA(anti-Proton Annihilation at
Darmstadt) Collaboration will study the $p\bar{p}$ reaction at FAIR,
with the $\bar{p}$ beam energy in the range of $1.5$ to $15$ $GeV/c$
and luminosity of about $10^{31}cm^{-2}s^{-1}$\cite{panda}. The
range of the beam energy is very suitable to find the $N^{*}(4265)$
and the $N^{*}(4418)$, with cross sections estimated to be about $70
n b$ and $2 n b$ by the one-meson exchange propagators with
off-shell form factors, which corresponds to an event production
rate of 60000 and 1700 per day at PANDA/FAIR, or about $10 n b$ and
$0.02 n b$ by the Reggeon propagators, which corresponds to an event
production rate of 9000 and 20 per day at PANDA/FAIR. There is a
$4\pi$ solid angle detector with good particle identification for
charged particles and photons at PANDA/FAIR. For the $p\bar{p}
\rightarrow p\bar{p} \eta_{c}$ reaction, if $p$ and $\bar{p}$ are
identified, then the $\eta_{c}$ can be easily reconstructed from the
missing mass spectrum against $p$ and $\bar{p}$. It is the same as
the reaction
 $p\bar{p} \rightarrow p\bar{p}
J/\psi$. So this reaction should be accessible at PANDA/FAIR.

\section{Summary}

In summary, we find six states from PB and VB channels by using the
local hidden gauge Lagrangian in combination with unitary techniques
in coupled channels. All of these states have large $c\bar{c}$
components, so their masses are all larger than 4200MeV. The width
of these states decaying to light meson and baryon channels without
$c\bar{c}$ components are all very small. On the other hand, the
$c\bar{c}$ meson - light baryon channels are also considered to
contribute to the decay width to these states. Then $\eta_{c}N$ and
$\eta_{c}\Lambda$ are added to the PB channels, while $J/\psi N$ and
$J/\psi \Lambda$ are added in the VB channels. The decay widths to
these channels are not negligible, in spite of the small phase space
for the decay, because the exchange $D^{*}(or D^{*}_{s})$ mesons
were less off-shell than the corresponding one in the decay to light
meson - light baryon channels. The total width of these states are
still very small. We made some estimates of cross sections for
production of these resonances at the upcoming FAIR facility. The
cross section of the reaction $p\bar{p} \rightarrow p\bar{p}
\eta_{c}$ and $p\bar{p} \rightarrow p\bar{p} J/\psi$ are about
$10\sim 70 n b$ and $0.02\sim 2 n b$, in which the main contribution
comes from the predicted $N^{*}_{c\bar{c}}(4265)$ and
$N^{*}_{c\bar{c}}(4418)$ states, respectively. With this theoretical
results, one can estimate about $9000\sim 60000$ and $20\sim 1700$
events per day at the PANDA/FAIR facility, respectively.

The predicted $N^{*}_{c\bar{c}}$ and $\Lambda^{*}_{c\bar{c}}$ can be
also looked for by many other processes, such as $ep\to e
N^{*}_{c\bar{c}}$ at JLab's 12 GeV upgrade, $Kp\to
\Lambda^{*}_{c\bar{c}}$ at JPARC, pp collisions, etc.

\section*{Acknowledgments}
We thank Li-sheng Geng and Feng-kun Guo for useful discussions. This
work is partly supported by DGICYT Contract No. FIS2006-03438, the
Generalitat Valenciana in the project PROMETEO, the EU Integrated
Infrastructure Initiative Hadron Physics Project under contract
RII3-CT-2004-506078, the National Natural Science Foundation of
China (NSFC) under grants Nos. 10875133, 10821063, 11035006, the
Chinese Academy of Sciences under project No. KJCX2-EW-N01, and the
Ministry of Science and Technology of China (2009CB825200).

\appendix
\section{The $C_{ab}$ coefficients}
\label{app:tables}

In this Appendix we give the coefficients $C_{ab}$ in
Eqs.~(\ref{vpbb}, \ref{vvbb}, \ref{widpb},\ref{pbetacb}) for the
several $(I,S)$ sectors studied in this work.
\begin{table}[H]
 \renewcommand{\arraystretch}{1.2}
\centering

\caption{ Coefficients $C_{ab}$ in the Eq.~(\ref{vpbb},
\ref{widpb}) for the $PB$ system in the sector $I=3/2$, $S=0$.}
\vspace{0.5cm}
\begin{tabular}{l|ccc}
 & $\bar{D} \Sigma_{c}$  & $\pi N$ & $K \Sigma$\\
 \hline
$\bar{D} \Sigma_{c}$ & $2$ & $-1$ & $1$\\
\end{tabular}
\end{table}
\begin{table}[H]
 \renewcommand{\arraystretch}{1.2}
\centering

\caption{ Coefficients $C_{ab}$ in the Eq.~(\ref{vpbb},
\ref{widpb}, \ref{pbetacb})  for the $PB$ system in the sector
$I=1/2$, $S=0$.} \vspace{0.5cm}
\begin{tabular}{l|cccccccc}
 & $\bar{D} \Sigma_{c}$ & $\bar{D} \Lambda^{+}_{c}$ & $\eta_{c} N$  & $\pi N$ & $\eta N$ & $\eta' N$ & $K \Sigma$ & $K \Lambda$\\
 \hline
$\bar{D} \Sigma_{c}$     & $-1$  &  $ 0$   & $-\sqrt{3/2}$  & $-1/2$  &   $-1/\sqrt{2}$    &   $1/2$   &     $1 $  &  $  0$          \\
$\bar{D} \Lambda^{+}_{c}$&       &  $ 1$ &  $\sqrt{3/2}$  & $-3/2$  &    $1/\sqrt{2}$   &  $-1/2$   &     $0$     &    1        \\
\end{tabular}
\end{table}
\begin{table}[H]
 \renewcommand{\arraystretch}{1.2}
\centering

\caption{
 Coefficients $C_{ab}$ in the Eq.~(\ref{vpbb}, \ref{widpb}) for the $PB$ system in the sector $I=1/2$, $S=-2$.} \vspace{0.5cm}
\begin{tabular}{l|cccccccc}
 & $\bar{D}_{s} \Xi^{'}_{c}$ &  $\bar{D}_{s} \Xi_{c}$ & $\bar{D} \Omega_{c}$  & $\pi \Xi$ &  $\bar{K} \Sigma$ & $\eta \Xi$ & $\eta' \Xi$ & $\bar{K} \Lambda$\\
 \hline
$\bar{D}_{s} \Xi^{'}_{c}$     & $1$  &   $0$     &   $\sqrt{2}$  & $0$              &   $\sqrt{3}/2$    &   $1/\sqrt{6}$  &   $1/\sqrt{3}$  &  $-\sqrt{3}/2$ \\
$\bar{D}_{s} \Xi_{c}$         &       &   $1$   &     $0$           & $0$              &  $-3/2$              &   $1/\sqrt{2}$   &  $ 1$          &   $1/2$\\
$\bar{D} \Omega_{c}$          &       &         &    $ 0$           & $\sqrt{3/2}$   &   $0$                &  $-1/\sqrt{3}$  &  $1/\sqrt{6}$  &   $0$   \\
\end{tabular}
\end{table}
\begin{table}[H]
 \renewcommand{\arraystretch}{1.2}
\centering

 \caption{
 Coefficients $C_{ab}$ in the Eq.~(\ref{vpbb}, \ref{widpb}) for the $PB$ system in the sector $I=1$, $S=-1$.} \vspace{0.5cm}
\begin{tabular}{l|ccccccccc}
 & $\bar{D}_{s} \Sigma_{c}$ &  $\bar{D} \Xi^{'}_{c}$ & $\bar{D} \Xi_{c}$  & $\pi \Sigma$  & $\pi \Lambda$ & $\eta \Sigma$ & $\eta' \Sigma$  & $\bar{K}N$  & K $\Xi$       \\
 \hline
$\bar{D}_{s} \Sigma_{c}$     & $0$     &   $\sqrt{2}$     &  $0$           &     $0$            &  $0$              &   $-1/\sqrt{3}$    &   $\sqrt{2/3}$   &  $-1$   & $0$\\
$\bar{D} \Xi^{'}_{c}$        &       &   $1$              & $0$       &  $1/\sqrt{2}$     &   $-\sqrt{3}/2$     &    $1/\sqrt{6}$    &   $1/2\sqrt{3}$  &   $0$    & $1/\sqrt{2}$\\
$\bar{D} \Xi_{c}$            &       &                    &   $1$         & $-\sqrt{3/2}$    &   $1/2$              &   $-1/\sqrt{2}$     &   $-1/2$           &   $0$      & $\sqrt{3/2}$\\
\end{tabular}
\end{table}
\begin{table}[H]
 \renewcommand{\arraystretch}{1.2}
\centering

\caption{
 Coefficients $C_{ab}$ in the Eq.~(\ref{vpbb}, \ref{widpb}, \ref{pbetacb}) for the $PB$ system in the sector $I=0$, $S=-1$.} \vspace{0.5cm}
\begin{tabular}{l|cccccccccc}
 & $\bar{D}_{s} \Lambda^{+}_{c}$ &  $\bar{D} \Xi_{c}$ & $\bar{D} \Xi^{'}_{c}$ & $\eta_{c}\Lambda$  & $\pi \Sigma$     &  $\eta \Lambda$    & $\eta' \Lambda$   & $\bar{K}N$     & K $\Xi$             \\
 \hline
$\bar{D}_{s} \Lambda^{+}_{c}$     & $0$     & $-\sqrt{2}$     &   $0$      & $1$                    &  $0$               &  $1/\sqrt{3}$     &  $\sqrt{2/3}$    &  $-\sqrt{3}$   & $0$\\
$\bar{D} \Xi_{c}$                 &       &  $-1$             &   $0$      &  $1/\sqrt{2}$         & $-3/2$             &  $1/\sqrt{6}$   & $-1/2\sqrt{3}$   &   0              & $\sqrt{3/2}$  \\
$\bar{D} \Xi^{'}_{c}$             &       &                    &  $-1$    & $-\sqrt{3/2}$         &  $\sqrt{3}/2$    & $-1/\sqrt{2}$    &  $1/2$             &   $0$              & $1/\sqrt{2}$\\
$\eta_{c}\Lambda$                 &       &                    &          & $0$                      &  $0$               & $ 0$                &  $0$               &   $0$              & $0$\\
\end{tabular}
\end{table}
\begin{table}[H]
 \renewcommand{\arraystretch}{1.2}
\centering

\caption{
 Coefficients $C_{ab}$ in the Eq.~(\ref{vpbb}, \ref{widpb}) for the $PB$ system in the sector $I=0$, $S=-3$.} \vspace{0.5cm}
\begin{tabular}{l|cc}
 & $\bar{D}_{s} \Sigma_{c}$ & $\bar{K}\Xi$\\
 \hline
$\bar{D}_{s} \Sigma_{c}$  & $2$  & $\sqrt{2}$  \\
\end{tabular}
\end{table}

\begin{table}[H]
 \renewcommand{\arraystretch}{1.2}
\centering

\caption{
 Coefficients $C_{ab}$ in the Eq.~(\ref{vvbb}, \ref{widpb}) for the $VB$ system in the sector $I=3/2$, $S=0$.}
\vspace{0.5cm}
\begin{tabular}{l|ccc}
 & $\bar{D}^{*} \Sigma_{c}$  & $\rho N$ & $K^{*} \Sigma$\\
 \hline
$\bar{D}^{*} \Sigma_{c}$ & $2$ & $-1$ & $1$\\
\end{tabular}
\end{table}

\begin{table}[H]
 \renewcommand{\arraystretch}{1.2}
\centering

\caption{
 Coefficients $C_{ab}$ in the Eq.~(\ref{vvbb}, \ref{widpb}) for the $VB$ system in the sector $I=1/2$, $S=0$.} \vspace{0.5cm}
\begin{tabular}{l|ccccccc}
 & $\bar{D}^{*} \Sigma_{c}$ & $\bar{D}^{*} \Lambda^{+}_{c}$ & $\rho N$ & $\omega N$ & $\phi N$ & $K^{*} \Sigma$ & $K^{*} \Lambda$\\
 \hline
$\bar{D}^{*} \Sigma_{c}$     & $-1$  &   $0 $   & $-1/2$ &    $\sqrt{3}/2$    &   $0$   &     $1$  &    $0$          \\
$\bar{D}^{*} \Lambda^{+}_{c}$&       &   $1 $ & $-3/2$  &   $-\sqrt{3}/2$    &   $0$   &     $0$     &    $1$        \\
\end{tabular}
\end{table}

\begin{table}[H]
 \renewcommand{\arraystretch}{1.2}
\centering

\caption{
 Coefficients $C_{ab}$ in the Eq.~(\ref{vvbb}, \ref{widpb}) for the $VB$ system in the sector $I=1/2$, $S=-2$.} \vspace{0.5cm}
\begin{tabular}{l|cccccccc}
 & $\bar{D}^{*}_{s} \Xi^{'}_{c}$ &  $\bar{D}^{*}_{s} \Xi_{c}$ & $\bar{D}^{*} \Omega_{c}$  & $\rho \Xi$ &  $\bar{K}^{*} \Sigma$ & $\omega \Xi$ & $\phi \Xi$ & $\bar{K}^{*} \Lambda$\\
 \hline
$\bar{D}^{*}_{s} \Xi^{'}_{c}$     &  $1$  &  $0$      &   $\sqrt{2}$  &  $0$              &   $\sqrt{3}/2$    & $0$              &   $-1/\sqrt{2}$  &  $-\sqrt{3}/2$ \\
$\bar{D}^{*}_{s} \Xi_{c}$         &       &   $1$     &       $0$     &   $0$             &  $-3/2$             &  $0$             &   $-\sqrt{3/2}$   &   $1/2$ \\
$\bar{D}^{*} \Omega_{c}$          &       &           &    $0$        & $\sqrt{3/2}$      &  $0$               &   $\sqrt{3}/2$   &   $0$               &   $0$    \\
\end{tabular}
\end{table}

\begin{table}[H]
 \renewcommand{\arraystretch}{1.2}
\centering

\caption{
 Coefficients $C_{ab}$ in the Eq.~(\ref{vvbb}, \ref{widpb}) for the $VB$ system in the sector $I=1$, $S=-1$.} \vspace{0.5cm}
\begin{tabular}{l|ccccccccc}
 & $\bar{D}^{*}_{s} \Sigma_{c}$ &  $\bar{D}^{*} \Xi^{'}_{c}$ & $\bar{D}^{*} \Xi_{c}$  & $\rho \Sigma$  & $\rho \Lambda$ & $\omega \Sigma$ & $\phi \Sigma$  & $\bar{K}^{*}N$ & $K^{*} \Xi$ \\
 \hline
$\bar{D}^{*}_{s} \Sigma_{c}$     & $0$     &   $\sqrt{2}$     &  $0$          &        $0$         &       $0$             & $0$                  &   $-1$          &  $-1$  & $0$          \\
$\bar{D}^{*} \Xi^{'}_{c}$        &       &   $1$             &   $0$           &  $1/\sqrt{2}$      &   $-\sqrt{3}/2$    &  $ -1/2$              &    $0$             &   $0$    &$1/\sqrt{2}$  \\
$\bar{D}^{*} \Xi_{c}$            &       &                    &   $1$         & $-\sqrt{3/2}$      &   $1/2$                &    $\sqrt{3}/2$     &    $0$             &   $0$    & $\sqrt{3/2}$  \\
\end{tabular}
\end{table}

\begin{table}[H]
 \renewcommand{\arraystretch}{1.2}
\centering \caption{
 Coefficients $C_{ab}$ in the Eq.~(\ref{vvbb}, \ref{widpb}) for the $VB$ system in the sector $I=0$, $S=-1$.} \vspace{0.5cm}
\begin{tabular}{l|cccccccc}
 & $\bar{D}^{*}_{s} \Lambda^{+}_{c}$ &  $\bar{D}^{*} \Xi_{c}$ & $\bar{D}^{*} \Xi^{'}_{c}$ & $\rho \Sigma$ &  $\omega \Lambda$ & $\phi \Lambda$ & $\bar{K}^{*}N$   & $K^{*} \Xi$ \\
 \hline
$\bar{D}^{*}_{s} \Lambda^{+}_{c}$     & $0$     &  $-\sqrt{2}$     &   $0$         &    $0$             & $0$                  &  $-1$             &  $-\sqrt{3}$     & $0$           \\
$\bar{D}^{*} \Xi_{c}$                 &       &  $-1$              &   $0$        &  $-3/2$            &   $-1/2$            &    $0$              &    $0$               & $\sqrt{3/2}$  \\
$\bar{D}^{*} \Xi^{'}_{c}$             &       &                    &  $-1$      & $\sqrt{3}$/2     &   $\sqrt{3}/2$    &   $0$              &   $0$           & $1/\sqrt{2}$  \\
\end{tabular}
\end{table}

\begin{table}[H]
 \renewcommand{\arraystretch}{1.2}
\centering
 \caption{
 Coefficients $C_{ab}$ in the Eq.~(\ref{vvbb}, \ref{widpb}) for the $VB$ system in the sector $I=0$, $S=-3$.} \vspace{0.5cm}
\begin{tabular}{l|cc}
& $\bar{D}^{*}_{s} \Sigma_{c}$ & $\bar{K}^{*}\Xi$\\
 \hline
$\bar{D}^{*}_{s} \Sigma_{c}$  & $2$  & $\sqrt{2}$ \\
\end{tabular}
\end{table}

\end{document}